\documentclass{aa}
\usepackage{graphicx}
\usepackage{longtable}
\usepackage{lscape}

\def\gsim{\;\lower.6ex\hbox{$\sim$}\kern-7.75pt\raise.65ex\hbox{$>$}\;}
\def\lsim{\;\lower.6ex\hbox{$\sim$}\kern-7.75pt\raise.65ex\hbox{$<$}\;}

\begin{document}
\title{Na-O Anticorrelation and HB. VI. The chemical composition
of the peculiar bulge globular cluster NGC~6388
\thanks{
Based on observations collected at ESO telescopes under programme 073.D-0211.
Tables 3 is only available in electronic form
at the CDS via anonymous ftp to cdsarc.u-strasbg.fr (130.79.128.5)
or via http://cdsweb.u-strasbg.fr/cgi-bin/qcat?J/A+A/}
 }

\author{
E. Carretta\inst{1},
A. Bragaglia\inst{1},
R.G. Gratton\inst{2},
Y. Momany\inst{2},
A. Recio-Blanco\inst{3},
S. Cassisi\inst{4},
P. Fran\c cois\inst{5,6},
G. James\inst{5,6},
S. Lucatello\inst{2},
\and
S. Moehler\inst{7}
}

\authorrunning{E. Carretta et al.}
\titlerunning{Chemical composition of NGC~6388}

\offprints{E. Carretta, eugenio.carretta@oabo.inaf.it}

\institute{
INAF-Osservatorio Astronomico di Bologna, Via Ranzani 1, I-40127
 Bologna, Italy
\and
INAF-Osservatorio Astronomico di Padova, Vicolo dell'Osservatorio 5, I-35122
 Padova, Italy
\and
Dpt. Cassiop\'ee, UMR 6202, Observatoire de la C\^ ote d'Azur, B.P. 4229, 06304
Nice Cedex 04, France
\and
INAF-Osservatorio Astronomico di Collurania, Via M. Maggini, I-64100 Teramo,
Italy
\and
Observatoire de Paris, 61 Avenue de l'Observatoire, F-75014 Paris, France
\and
European Southern Observatory, Alonso de Cordova 3107, Vitacura, Santiago,Chile
\and
European Southern Observatory, Karl-Schwarzchild-Strasse 2, D-85748 Garching
bei Munchen, Germany
  }

\date{8 Jan 2007}

\abstract{
We present the LTE abundance analysis of high resolution spectra for red giant
stars in the peculiar bulge globular cluster NGC~6388. Spectra of seven members
were taken using the UVES spectrograph at the ESO VLT2 and the multiobject 
FLAMES facility. We exclude any intrinsic metallicity spread in this cluster:
on average, [Fe/H]$=-0.44\pm0.01\pm0.03$ dex on the scale of the present series
of papers, where the first error bar refers to individual star-to-star errors
and the second is systematic, relative to the cluster. Elements involved in
H-burning at high temperatures show large spreads, exceeding the estimated
errors in the analysis. In particular, the pairs Na and O, Al and Mg
are anticorrelated and Na and Al are correlated among the giants in NGC~6388,
the typical pattern observed in all galactic globular clusters studied so far.
Stars in NGC~6388 shows an excess of $\alpha-$process elements, similar to the
one found in the twin bulge cluster NGC~6441. Mn is found underabundant in
NGC~6388, in agreement with the average abundance ratio shown by clusters of
any metallicity. Abundances of neutron-capture elements are homogeneously
distributed within NGC~6388; the [Eu/Fe] ratio stands above the value found
in field stars of similar metallicity.
\keywords{Stars: abundances -- Stars: atmospheres --
Stars: Population II -- Galaxy: globular clusters -- Galaxy: globular
clusters: individual: NGC~6388}} 

\maketitle

\section{Introduction}

NGC~6388 (C1732-447) is one of the 10 more massive objects in the whole
globular cluster (GC) system of our Galaxy.
Located in the bulge like its cluster twin NGC~6441, NGC~6388 is a luminous
object ($M_V=-9.42$ according to the updated on-line version of the Harris 1996
catalog) at about 3 kpc from the Galactic center, at
10 kpc from the Sun and at slightly more than 1 kpc from the Galactic plane.
Unfortunately, nothing is known about its orbit, due to the lack of
any proper motion informations, or about its age.

Until about 10 years ago, this cluster was considered nothing more than another
moderately metal-rich ([Fe/H]$=-0.60 \pm 0.11$ dex, Armandroff and Zinn 1988,
from integrated light spectroscopy) bulge cluster.
However, the discovery by Rich et al. (1997) of an extended blue
horizontal-branch (HB) in NGC~6388 (as well as in NGC~6441) renewed 
interest and raised many questions about this pair of clusters. 
In case of (likely) old, metal-rich GCs we would expect from stellar evolution 
theory that HB stars were concentrated only to the red of the instability strip.
NGC~6388 (and NGC~6441) are living proofs of something unusual, at
variance with "well behaved" clusters such as 47 Tuc and the near-solar 
metallicity Galactic open clusters, with their red HBs (clumps).

Besides the red HB clump, these two GCs show a conspicuous, well developed 
blue HB as well as a substantial  population of RR Lyrae variables (Pritzl et
al. 2002, Corwin et al. 2006); hence they qualify as the most metal rich
globular cluster able to host Pop II variable stars (RR Lyraes as well as Pop
II Cepheids). 
It is clear that the HB morphology and its variation with metallicity in this
pair of clusters  has no counterpart among the other Galactic GCs.

NGC~6388 (together with NGC~6441) is the most metal-rich example of the
second-parameter effect at work in modeling the HB morphology (the first 
parameter being the metal abundance). Moreover, it seems that
whatever the effect(s) is (are), the consequences are visible within the same
cluster. This phenomenon is not unknown, the most notable example
being NGC 2808, with its famous bimodal distribution of stars into a red HB
clump and a blue HB extending down to faint magnitudes (e.g. Momany et al.
2003). However, important differences do exist: (i) NGC 2808 has a lower
metallicity ([Fe/H]$=-1.14$, Carretta 2006), however, only a scarce population
of variable stars has been found (Corwin et al. 2004); instead, both NGC~6388 and
NGC~6441 host a relevant number of stars within the instability strip (Pritzl
et al. 2001, 2002); (ii) the blue HB
in NGC 2808 is normal, beginning at the same level of magnitude of the redder 
part; on the other hand, both bulge clusters have a blue HB sloped upward
going blue ward in the $V, B-V$ color-magnitude diagram (Raimondo et al. 2002).
In other words, the top of the blue HB tail is about 0.5 mag brighter in $V$
than the red HB clump; this tilt is not due to the differential reddening
affecting NGC~6388 (see discussion in Raimondo et al. 2002) and it is
present in all the photometric pass bands (Busso et al. 2004). 
Moreover, this is
consistent with RR Lyraes in these extreme second-parameter clusters having
abnormally long periods for their high metallicities, in fact even longer
than in more metal-poor, typical Oosterhoff II GCs. 

One of the most favored explanations for these peculiarities involves He
enrichment in dense environments. Very blue, extended HBs may be a consequence
of distinct stellar populations characterized by distinct He abundances, as
suggested by D'Antona and Caloi (2004, and references therein). Also 
scenarios involving either a cluster high primordial He abundance or 
an He enhancement due to deep mixing have been proposed (see Sweigart and
Catelan 1998). However, the latter solutions appear to be ruled out, and recent
spectroscopic analyses of hot HB stars in NGC~6388 support the He pollution
scenario (Moehler and Sweigart 2006a,b).

Therefore, NGC~6388 was well entitled to be
considered (together with NGC~6441) in our present project aimed to study in
details the possible link between chemical anomalies and global parameters (in 
particular the HB morphology). In this effort, we used the FLAMES multiobject
facility (Pasquini et al. 2002) to obtain extensive spectroscopic surveys of 
about 100 stars on the red giant branch (RGB) in about 20 GCs. The observed
spectra are used to derive 
abundances of Na and O, the best known elements involved in the high 
temperature H-burning reactions. In turn, the well known anticorrelation
between Na and O abundances in RGB stars (see Gratton et al. 2004 for an
extensive review on this subject) should be in some way linked to the existence
of a He-rich population. In fact, the observed Na enhancements and O depletions 
(the typical signature of matter processed in intermediate-mass AGB stars and
ejected in the early cluster environment) are expected to be accompanied by
alterations in the He content. Thus, the evolving stars formed out of
polluted gas should end up onto the bluest part of the HB, during the phase of 
their He-core burning since they evolve faster on the main sequence. All other
factors (including the mass loss rate along the RGB) being equal, they will start
their HB phase at hotter temperature, being less massive than other normal
stars at the end of the main sequence (see D'Antona and Caloi 2004 and
references therein).

In the framework of the present project we have already examined the massive
and peculiar cluster NGC~2808 (Carretta et al. 2006a; Paper I),
the more "normal", intermediate-metallicity, blue HB cluster NGC~6752 
(Carretta et al. 2006b, hereafter Paper II), the other anomalous bulge cluster 
NGC~6441 (Gratton et al. 2006, 2007 Paper III and Paper V) and NGC 6218 
(Carretta et al. 2006c, Paper IV).

The Na-O anticorrelation in NGC~6388, based on the detailed analysis of the
GIRAFFE data will be presented in a forthcoming paper, while we concentrate
here on the analysis of the higher resolution UVES spectra.
Deciphering the detailed chemical composition of
NGC~6388 is of paramount importance, because of
the peculiar features of this cluster (and of its associate NGC~6441):
\begin{itemize}
\item[(a)] these clusters are among the best examples available to us of old,
metal-rich populations, hence a key ingredient in the interpretation and
synthesis of distant elliptical galaxies; 
\item[(b)] the presence of an important population of hot
HB stars, presently under detailed investigations (Moehler et al. 1999, Moehler
and Sweigart 2006b), is considered the major contributor to the UV-upturn
phenomenon (see Yi et al. 1998 and references therein); 
\item [(c)] deriving the
precise metal abundance of these clusters is a key task in establishing the
universality of the $M_{\rm V}$ vs [Fe/H] relation (or, more dramatically, its
violation) for RR Lyraes, one of the best Population II standard candles
(Sandage 2006); 
\item[(d)] photometric evidences already show that a large metallicity spread 
is not consistent with the CMD features in both these clusters (Raimondo et al.
2002); 
previous spectroscopic analyses in NGC~6441 (high resolution spectra of giants,
Paper III; low resolution spectroscopy of RR Lyraes, Clementini et al.
2006) and in NGC~6388 (high-resolution spectra of hot HB stars, Moehler and
Sweigart 2006a) also show no evidence for any deviation in metallicity from the
cluster. Nevertheless, it is necessary to assess with homogeneous analysis 
if a real metallicity spread can be firmly excluded also in NGC~6388;
\item[(e)] finally, both NGC~6388 and NGC~6441 have the highest predicted
escape velocity at cluster center among 153 resolved star clusters in the
Milky Way, the Magellanic Clouds and the Fornax dSph galaxy (McLaughlin and Van
der Marel 2006), more than 3$\sigma$ away of the average velocity of the other
Galactic GCs, including the most massive ones. This implies that they
are very tightly bound (as they must, having survived almost an Hubble time
likely very close to the bulge of the Galaxy). In turn, we could have the
chance of observing a 
clear signature of the predicted (e.g. Cayrel 1986) and long searched for,
self-enrichment: perhaps these GCs were able to efficiently retain at least part of
the metal-enriched ejecta of core-collapse Supernovae. However, we must note
that a deep search for neutral hydrogen emission at 21 cm (Lynch et al. 1989),
only resulted in an upper limit of $M_{H {\sc i}} = 3.8 M_\odot$.
\end{itemize}

Thus, as done with the companion cluster NGC~6441 (Paper III), we
used the high resolution spectra provided by the fiber-fed UVES spectrograph,
exploiting the large spectral coverage to derive the detailed chemical
composition of NGC~6388.

The present paper is organized as follows: an outline of the observations 
and target selection is given in Sect. 2; the analysis, including the
derivation of the atmospheric parameters and metallicities, is described in
Sect. 3, whereas in Sect. 4 we discuss the error budget of our abundance
analysis. Abundances of various elements are presented in Sect. 5; finally, 
Sect. 6 provides a summary of the study.

\section{Target selection, observations and membership}

Our data were collected in Service mode 
with the ESO high resolution multifiber spectrograph 
FLAMES (Pasquini et al. 2002) mounted on VLT. The UVES Red Arm provided a 
wavelength coverage from 4800 to 6800~\AA\, with a resolution R$\simeq 40,000$.
The observations 
required to reach the necessary S/N were done over a period of several nights.

We used two fiber  configurations in order to maximize the number of possible
targets  observed with the dedicated fibers feeding the UVES spectrograph. 
In the first configuration (8 exposures), seven stars were
observed, with one fiber pointing to the sky; in the second set (10 exposures),
we observed six stars, with two fibers dedicated to the sky. 
In Table~\ref{t:log} we list all the relevant informations.

The target selection followed the same criteria of the previous GCs;
we chose stars near
the RGB ridge line and isolated\footnote{All stars were chosen to be free from
any companion brighter than $V+2$ mag within a
2.5~arcsec radius, or brighter than $V-2$ mag within 10~arcsec, where $V$ is
the target magnitude.}. No previous studies about membership (either 
based on spectroscopy or on proper motions) were available. Moreover, 
the constraints imposed by mechanical
limitations in the Oz-Poz fiber positioner for FLAMES and the size  on the sky
of this rather concentrated cluster forced us to observe stars at some distance
from the center, increasing the risk to include field interlopers.

\begin{figure}
\centering
\includegraphics[bb=20 150 580 710, clip, scale=0.45]{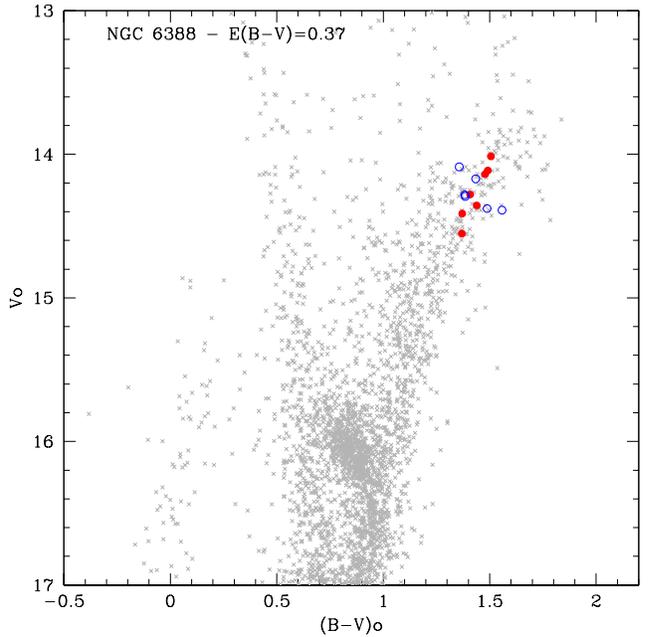}
\caption{Dereddened $V,B-V$ color-magnitude diagram for NGC~6388 in a 
5$\farcm$0 selection box around the cluster's center. WFI photometry is from
Momany et al. (2003); only stars  with absolute {\it SHARP}  values $\le0.8$
are plotted. Stars observed with UVES and members of NGC~6388 according to
their RVs are indicated by (red) filled circles; open circles indicates non
member stars observed with UVES.}
\label{f:cmd63}
\end{figure}

Another problem in the selection is due to
the differential reddening affecting NGC~6388, which spreads the width of the
RGB. The combination of technical limitations in the fiber positioning, spread
of sequences and limited number of bright stars made difficult to restrict the
observed sample only to objects very close to the RGB ridge line. 
Stars observed with UVES are indicated by large circles in
Figure~\ref{f:cmd63}, whereas in Table~\ref{t:coou63} all targets are listed.
Star designations, optical magnitudes and coordinates (J2000
equinox) are from Momany et al. (2003); heliocentric radial velocities are
those derived from our spectra (see below).

\begin{table}\centering
\caption{Log of the observations for NGC~6388. Date and time are UT, exposure
times are in seconds. For both fiber configurations the field center is at
RA(2000)=17:35:59, Dec(2000)=$-$44:42:32}
\begin{tabular}{ccccc}
\hline
Fiber        &Date       &UT$_{beginning}$ &exptime &airmass \\
config.&           &                 &        &        \\
\hline
1 & 2004-05-04 & 04:19:33 & 4000 & 1.36 \\
1 & 2004-05-04 & 05:29:35 & 4200 & 1.17 \\
1 & 2004-05-04 & 06:50:30 & 4000 & 1.07 \\
1 & 2004-05-11 & 08:51:04 & 4000 & 1.15 \\
1 & 2004-05-11 & 08:02:24 & 2700 & 1.09 \\
1 & 2004-05-12 & 05:12:58 & 4500 & 1.14 \\
1 & 2004-06-28 & 04:38:20 & 4000 & 1.08 \\
1 & 2004-07-11 & 00:43:11 & 4000 & 1.21 \\
2 & 2004-07-21 & 00:21:10 & 4200 & 1.07 \\
2 & 2004-07-21 & 02:33:15 & 4000 & 1.09 \\
2 & 2004-07-21 & 03:42:17 & 4000 & 1.11 \\
2 & 2004-07-25 & 00:28:49 & 4000 & 1.13 \\
2 & 2004-07-25 & 01:40:20 & 4200 & 1.17 \\
2 & 2004-07-25 & 03:23:53 & 4000 & 1.11 \\
2 & 2004-07-25 & 04:32:19 & 4000 & 1.36 \\
2 & 2004-07-26 & 00:14:10 & 4000 & 1.15 \\
2 & 2004-07-26 & 01:23:07 & 4000 & 1.08 \\
2 & 2004-07-26 & 02:41:49 & 2700 & 1.08 \\
\hline
\end{tabular}
\label{t:log}
\end{table}

\paragraph{Photometry and astrometry}

We used high quality $BVI$ photometry obtained with the Wide Field Imager at
the 2.2 m ESO/MPI telescope in La Silla (ESO Programme 69.D-0582). The
optical camera WFI is a mosaic of eight CCD chips, each with  a field of view 
of ${\sim}8'{\times}16'$, for a total field of view of ${\sim}34'{\times}33'$. 
The cluster center was roughly centered on  chip~\#7, and the  images were
taken  through the $B$, $V$ and  $I_c$ broad band filters  with typical
exposure times of 120~sec, 90~sec and 50~sec, respectively.  The average seeing
was good and stable (full width half maximum $1\farcs0\le$FWHM$\le 1\farcs3$).

The photometric reduction  and calibration methods are well-tested and
have   been   presented  in Momany et al. (2002).  In
particular, NGC~6388 belongs to the  69.D-0582 observing run from which
wide field photometry for  other globular clusters have been published
(Momany  et   al. 2003).  In the   following  we  provide a
description of the photometric and astrometric calibration.
Stellar photometry  was performed using the standard
DAOPHOT~II and ALLFRAME  programs  (Stetson 1994).  In  crowded fields
such as that surrounding NGC~6388, these programs provide high quality
stellar photometry by point spread  function (PSF) fitting.  The single PSFs  
were constructed for each image
after a careful selection of isolated and well-distributed  stars
across each chip.

The  instrumental PSF magnitudes   of NGC~6388 were  normalized to 1~s
exposure and zero airmass; they were further corrected for the 
aperture correction, 
necessary  to  convert the instrumental
PSF magnitudes of NGC~6388  to the scale of the  standard
star  measurements (based  on  aperture photometry,  see below).   The
instrumental NGC~6388 magnitudes are therefore:

\begin{equation}
m^{'}=m_{\rm apert.}+2.5\log(t_{exp})-\kappa_{\lambda}X
\end{equation}

\noindent
where $m_{\rm apert.}=m_{\rm PSF}-{\rm apert.~correction}$, $X$ is the airmass
of the  reference image for  each filter, and the adopted mean extinction
coefficients for La  Silla are: $\kappa_B=0.23$, $\kappa_V=0.16$  and
$\kappa_I=0.07$.

The aperture  corrections  were estimated on isolated bright stars,
evenly  distributed   across  the chip.   For   these we obtained
aperture   photometry after   subtracting  any  nearby companions
within 5 FWHMs.  The aperture magnitudes measured in circular apertures
of radius  6 arcsecond  (close    to the photoelectric  aperture
employed by  Landolt 1992) were then  compared to  the PSF magnitudes,
and the difference is  assumed to be the  aperture correction to apply
to the PSF magnitudes.

Separately, aperture magnitudes of  standard stars from 3 Landolt
(1992) fields  were measured in  circular apertures of 6 arcsecond in
radius.  These  aperture   magnitudes    were  normalized   to   their
corresponding airmasses and    exposure times, and compared with   those
tabulated in  Landolt (1992). A  least  square fitting procedure provided
the calibration relations.

The $r.m.s.$ scatter of the residuals of the fit ($0.011$, $0.009$ and
$0.011$)  are assumed to  represent  our calibration uncertainties  in
$B$, $V$ and $I$ respectively.

The total zero-point uncertainties, including the aperture corrections
and calibration errors, are $0.015$, $0.013$,  and $0.016$ mag in $B$,
$V$ and $I$ respectively.  The  PSF photometry of NGC~6388 (normalized
and  corrected for aperture correction) was  then calibrated using the
relations:

\begin{equation}
B=b^{'}+0.287(B-V)+24.590
\end{equation}

\begin{equation}
V=v^{'}-0.078(B-V)+24.194 
\end{equation}

\begin{equation}
V=v^{'}-0.077(V-I)+24.199 
\end{equation}

\begin{equation}
I=i^{'}+0.097(V-I)+23.142
\end{equation}

The {\it USNO  CCD Astrograph all-sky  (UCAC2)}  Catalog (Zacharias et
al. 2004) was  used to search  for astrometric standards in the entire
WFI   image field of view.   Several  hundred astrometric {\it  UCAC2}
reference   stars were found in  each   chip, allowing an  accurate
absolute positioning of the  sources. An astrometric solution has been
obtained for   each of the eight   WFI  chips independently,  by using
suitable catalog matching  and cross\--correlation {\sc IRAF}\footnote{
IRAF is distributed by the National Optical Astronomical
Observatory, which are operated by the Association of Universities for
Research in Astronomy, under contract with the National Science
Foundation } tools. At  the
end   of  the  entire procedure,    the     {\it rms}  residuals   are
${\leq}$$0\farcs$2 both in right ascension and declination.

\begin{table*}     
\centering
\caption{List and relevant informations for the target stars observed
in NGC~6388 with UVES. ID, $B$, $V$ $I$  magnitudes and coordinates (J2000) are
those presented in this paper; $J$, $K$ magnitudes are from the 2MASS
catalog; the RVs (in km s$^{-1}$) are heliocentric. 
Members of NGC~6388 are  indicated by M in Notes, whereas stars non
members based on the RVs are indicated  by NM.  The  number in col. 2
indicates the  fiber configuration.}  
\begin{tabular}{rcccccccccrl}
\hline 
\hline Star  & conf.  &RA     &Dec   &$V$     &$B$     &$I$     &$J$    &$K$   &RV,hel.   &Notes \\ 
\hline
 77599&1& 17 35 53.602 &-44 48 15.84 &15.653  &17.424  & 13.805 &12.496 &11.415&   +76.09 & M  \\
 78569&1& 17 35 24.770 &-44 47 18.58 &15.328  &17.085  & 13.438 &12.188 &11.153& $-$91.30 & NM \\
 82375&1& 17 35 21.046 &-44 43 41.58 &15.522  &17.305  & 13.595 &12.324 &11.291& $-$12.07 & NM \\
 83168&2& 17 35 53.715 &-44 42 56.02 &15.793  &17.562  & 13.948 &12.581 &11.541&   +79.26 & M  \\
 85915&2& 17 35 49.560 &-44 39 59.65 &15.618  &17.506  & 13.355 &11.875 &10.650& $-$44.61 & NM \\
 86759&2& 17 35 28.083 &-44 39 04.95 &15.629  &17.587  & 12.972 &11.356 &10.090&$-$143.82 & NM \\
101131&1& 17 36 02.227 &-44 44 32.75 &15.379  &17.257  & 13.322 &11.877 &10.608&   +74.30 & M  \\
108176&2& 17 36 12.797 &-44 41 50.60 &15.254  &17.160  & 12.975 &11.472 &10.223&   +79.28 & M  \\
108895&1& 17 36 15.184 &-44 41 27.20 &15.596  &17.435  & 13.527 &12.062 &10.918&   +80.19 & M  \\
110677&1& 17 36 28.549 &-44 40 09.49 &15.520  &17.328  & 13.518 &12.185 &10.962&   +82.68 & M  \\
110754&2& 17 36 30.374 &-44 40 04.60 &15.533  &17.318  & 13.479 &12.113 &10.971&$-$156.28 & NM \\
111408&1& 17 36 06.921 &-44 39 29.10 &15.353  &17.244  & 13.204 &11.751 &10.519&   +82.10 & M  \\
111970&2& 17 36 04.630 &-44 38 54.64 &15.411  &17.245  & 13.254 &11.825 &10.480& $-$54.21 & NM \\
\hline
\end{tabular}
\label{t:coou63}
\end{table*}

\paragraph{Preparation of spectra and membership}

The spectra were reduced to 1-d, wavelength calibrated files by the ESO Service
mode personnel using the UVES-FLAMES pipeline (uves/2.1.1 version). 

We measured radial velocities (RVs) for each individual spectrum using about
80 atomic lines with the {\sc IRAF} package {\sc Rvidlines}. 
The resulting heliocentric RVs are shown in Table~\ref{t:coou63}; the
associated errors are typically a few hundreds of m/s.

The spectra were then shifted to zero radial velocity and co-added for each 
star. Final S/N values of the combined exposures range from 40 to 80 per pixel 
and were estimated in the spectral region near 5000~\AA. 

At variance with the case of NGC~6441 (Paper III), the determination of membership 
for stars in NGC~6388 was relatively
straightforward. This
cluster has an average radial velocity of +81.2$\pm 1.2$ km s$^{-1}$ (Harris
1996), and it lies in a region of the Galactic bulge where the bulk of
contaminating field stars have (according to our extensive spectroscopic
observations) a negative radial velocity.

\begin{figure}
\centering
\includegraphics[bb=20 150 580 710, clip, scale=0.45]{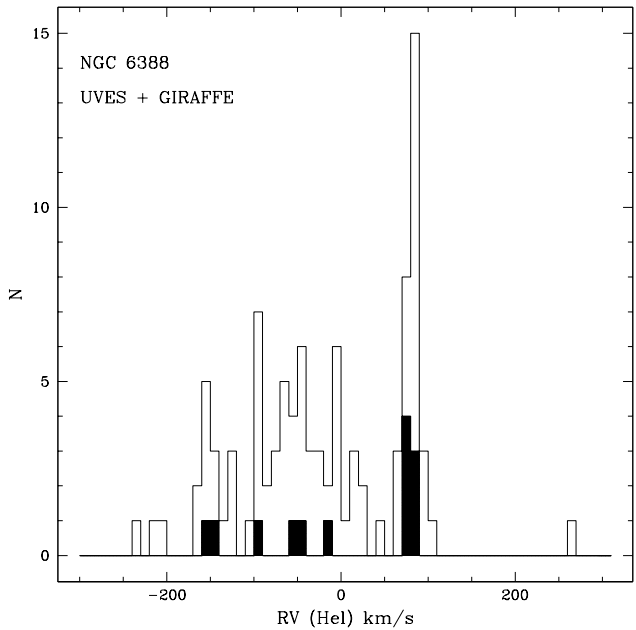}
\caption{Distribution of heliocentric RVs of all stars observed with UVES
(shaded area) in the present study as well as with GIRAFFE in Carretta et al.
(2006d).}
\label{f:rv6388}
\end{figure}

In Figure~\ref{f:rv6388} we show the heliocentric RV distribution of all the
stars observed with UVES (shaded area) in the present study and with GIRAFFE,
to investigate on the Na-O anticorrelation in NGC~6388 (Carretta et al. 2006d,
in preparation).
Stars in NGC~6388 stand out as a very neatly defined peak, whereas field stars
are recognizable as a broad distribution peaked at about -50 km s$^{-1}$; 
therefore, we could easily discriminate the true cluster members. The membership of
our program stars with UVES spectra is indicated in Table~\ref{t:coou63}: we
observed five cluster stars in the first fiber configuration and only two in
the second configuration. The non member stars were disregarded from further
analysis.
The average heliocentric RV derived from the final
sample of seven stars in NGC~6388 is +79.1$\pm 1.1$ (rms=3.0) km s$^{-1}$, in
good agreement with the tabulated value of Harris (1996).

\section{Atmospheric parameters, analysis and metallicity}

\subsection{Atmospheric parameters}

Temperatures and gravities were derived following the tested procedure adopted
in previous papers of this series.
We started obtaining $T_{\rm eff}$'s and bolometric corrections B.C. for 
our seven stars in NGC~6388 from
$V-K$ colors; these were the combination of our $V$ magnitudes and the 
$K$ magnitudes from the 2MASS Point Source
Catalogue (Skrutskie et al. 2006), transformed to
the TCS photometric system, as used in Alonso et al. (1999).
We employed the relations by Alonso et al.
(1999, with the erratum of 2001). We adopted for NGC~6388 a
distance modulus of $(m-M)_V$=16.14, a reddening of $E(B-V)$ = 0.37,  an input
metallicity  of [Fe/H]$=-0.60$  
(Harris 1996), and the relations  $E(V-K) = 2.75 E(B-V)$, $A_V =
3.1 E(B-V)$, and $A_K = 0.353 E(B-V)$ (Cardelli et al. 1989). 

However, our finally adopted $T_{\rm eff}$'s were
derived using a calibration between $T_{\rm eff}$ from dereddened $V-K$ colors 
and $K$ magnitudes based on 33 cluster star members (Carretta et al. 2006d).
As discussed in Paper II and Paper V, this procedure was adopted in order to 
decrease the scatter in abundances
due to uncertainties in temperatures, since magnitudes are much more reliably
measured than colors. The choice of $K$ magnitudes is particularly well suited
for GCs where a significant differential reddening is present (cfr. Paper III
and IV).

Surface gravities log $g$'s were obtained from the apparent magnitudes, the 
above effective temperatures and distance modulus, and the 
B.C's from Alonso et al. (1999), assuming  masses of 
0.90 M$_\odot$ and $M_{\rm bol,\odot} = 4.75$ as the bolometric magnitude 
for the Sun.
Along  with the derived atmospheric parameters (model metal abundance and
microturbulent velocities) and iron abundances, they are shown in Table
\ref{t:atmparu63}.

\subsection{Equivalent widths and iron abundances}

Line lists, atomic parameters and reference solar abundances are from Gratton
et al. (2003). We used the same procedure of previous papers, described in
details in Bragaglia et al. (2001), to measure equivalent widths ($EW$s) with
the ROSA code (Gratton 1988; see Table~\ref{t:tabewu63}, only available in
electronic form).

\begin{table*}
\centering
\caption[]{Equivalent widths from UVES spectra for stars of NGC 6388 (in
electronic form only).}
\begin{tabular}{lccrrrrrrrr}
\hline
El.         &$\lambda$&E.P.&$\log gf$&77599 & 83168 &101131 &108176 &108895 &110677 &111408 \\
            &         &    &         & EW   &  EW   &  EW   &  EW   &  EW   & EW    &  EW   \\
            & (\AA)   &(eV)&         &(m\AA)&(m\AA) &(m\AA) &(m\AA) &(m\AA) &(m\AA) &(m\AA) \\
\hline
O {\sc i}   &6300.31 & 0.00&  -9.75&   31.6 &  36.9 &  30.8 &  50.4 &  35.9 &  58.2 &  56.0 \\
O {\sc i}   &6363.79 & 0.02& -10.25&   16.3 &  20.4 &  12.2 &  29.0 &  14.5 &  27.6 &  34.1 \\
Na {\sc i}  &5682.65 & 2.10&  -0.67&  192.4 & 185.1 & 211.0 & 189.7 & 191.3 &	0.0 & 190.6 \\
Na {\sc i}  &5688.22 & 2.10&  -0.37&  188.0 & 179.4 & 204.7 &	0.0 & 184.4 & 175.9 & 192.2 \\
Na {\sc i}  &6154.23 & 2.10&  -1.57&  145.0 & 125.0 & 146.0 & 128.7 & 113.0 & 108.0 & 139.1 \\
Na {\sc i}  &6160.75 & 2.10&  -1.26&  163.0 & 136.0 & 154.2 & 138.0 & 133.2 & 125.0 & 145.8 \\
Mg {\sc i}  &5528.42 & 4.34&  -0.52&  249.8 & 238.6 & 249.0 & 250.1 & 256.3 & 243.5 & 247.7 \\
Mg {\sc i}  &5711.09 & 4.34&  -1.73&  139.4 & 146.4 &	0.0 & 139.2 &	0.0 &	0.0 &	0.0 \\
\hline
\end{tabular}
\label{t:tabewu63}
\end{table*}

Final metallicities are obtained with standard line analysis from measured
$EW$s and interpolating within the Kurucz (1993) grid of model
atmospheres (with the option for overshooting on) the model with the proper
atmospheric parameters whose abundance matches that derived from Fe {\sc i} 
lines.

Microturbulence velocities $v_t$ were derived by eliminating trends between
abundances from neutral Fe lines and expected line strength (see Magain 1984);
with this approach we avoid introducing spurious (positive) trends due to
systematic errors associated to the measure of observed $EW$s.

Average abundances of iron for NGC~6388 are [Fe/H]{\sc i}=$-0.44$ (rms=0.04 
dex) and [Fe/H]{\sc ii}=$-0.37$ (rms=0.09 dex) from 7 stars.
This difference (on average 0.07 dex with an rms=0.10 dex) 
is scarcely significant, and the good agreement between
abundances from neutral and singly ionized Fe lines indicates a 
self-consistent analysis. 

\begin{figure}
\centering
\includegraphics[clip, scale=0.45]{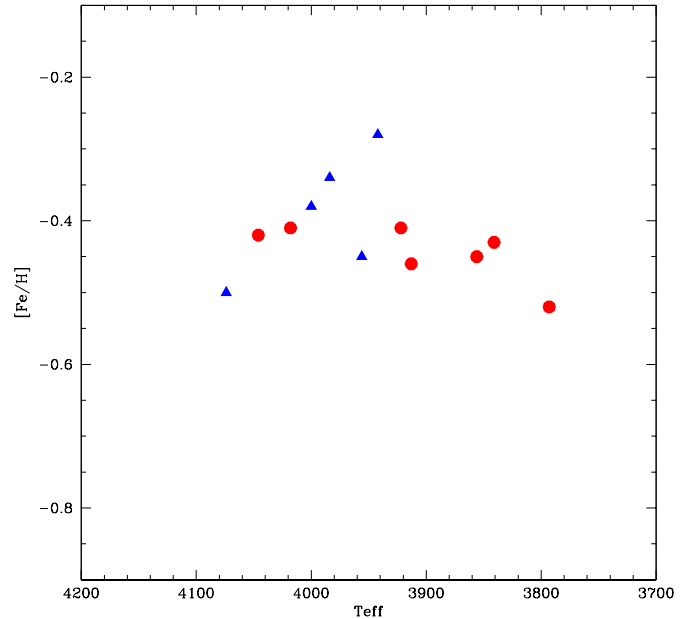}
\caption{Run of [Fe/H] ratios as a
function of temperatures for program stars in NGC~6388 (filled circles). 
As a comparison, stars observed with UVES in NGC~6441 from Gratton et al. 
(2006) are also shown (filled triangles).}
\label{f:feteffu63}
\end{figure}

There is no significant trend of iron abundance as a function of the
temperatures, as shown in Figure~\ref{f:feteffu63}, where we display the 
[Fe/H] ratios for stars observed with UVES in
NGC~6388 and, as a comparison, in NGC~6441 (Paper III).

The (small) scatter of the metallicity distribution is discussed below.

\begin{table*}
\centering
\caption[]{Adopted atmospheric parameters and derived iron abundances in 
stars of NGC~6388 from UVES spectra; nr indicates the number of lines 
used in the analysis. }
\begin{tabular}{rccccrcrccrcc}
\hline
Star   &  T$_{\rm eff}$ & $\log$ $g$ & [A/H]  &$v_t$	     & nr & [Fe/H]{\sc i} & $rms$ & nr & [Fe/H{\sc ii} & $rms$ \\
       &     (K)	&  (dex)     & (dex)  &(km s$^{-1}$) &    & (dex)	  &	  &    & (dex)         &       \\
\hline
 77599     & 4018  & 1.37 &$-$0.40 & 1.66 &  79  &$-$0.41 & 0.16 &  6  &$-$0.36 & 0.15 \\ 
 83168     & 4046  & 1.42 &$-$0.41 & 1.75 &  97  &$-$0.42 & 0.16 &  9  &$-$0.43 & 0.15 \\ 
101131     & 3856  & 1.02 &$-$0.46 & 1.68 &  89  &$-$0.45 & 0.17 &  8  &$-$0.53 & 0.26 \\ 
108176     & 3793  & 0.85 &$-$0.52 & 1.50 &  79  &$-$0.51 & 0.18 &  7  &$-$0.28 & 0.13 \\ 
108895     & 3913  & 1.15 &$-$0.45 & 1.65 & 100  &$-$0.46 & 0.17 & 10  &$-$0.33 & 0.18 \\ 
110677     & 3922  & 1.17 &$-$0.42 & 1.57 &  90  &$-$0.41 & 0.18 &  9  &$-$0.29 & 0.15 \\ 
111408     & 3841  & 0.98 &$-$0.43 & 1.57 &  91  &$-$0.42 & 0.19 &  7  &$-$0.36 & 0.07 \\ 
\hline
\end{tabular}
\label{t:atmparu63}
\end{table*}

\begin{table*}
\centering
\caption[]{Sensitivities of abundance ratios to variations in the atmospheric
parameters and to errors in the equivalent widths, and errors in abundances for
individual stars in NGC~6388 observed with UVES}
\begin{tabular}{lrrrrrrrr}
\hline
Element     & Average  & T$_{\rm eff}$ & $\log g$ & [A/H]   & $v_t$    & EWs     & Total   & Total      \\
            & n. lines &      (K)      &  (dex)   & (dex)   &kms$^{-1}$& (dex)   &Internal & Systematic \\
\hline        
Variation&             &  50           &   0.20   &  0.10   &  0.10    &         &         &            \\
Internal &             &   6           &   0.04   &  0.00   &  0.07    & 0.170   &         &            \\
Systematic&            &  57           &   0.06   &  0.01   &  0.03    & 0.170   &         &            \\
\hline
$[$Fe/H$]${\sc  i}& 89 &$-$0.009       &   +0.039 &  +0.027 & $-$0.049 & +0.018  &0.040    &0.028      \\
$[$Fe/H$]${\sc ii}&  8 &$-$0.101       &   +0.122 &  +0.044 & $-$0.036 & +0.060  &0.071    &0.135      \\
$[$O/Fe$]${\sc  i}&  2 &  +0.025       &   +0.042 &  +0.013 &   +0.046 & +0.120  &0.125    &0.125      \\
$[$Na/Fe$]${\sc i}&  4 &  +0.054       & $-$0.065 &$-$0.019 &   +0.011 & +0.085  &0.087    &0.107      \\
$[$Mg/Fe$]${\sc i}&  3 &$-$0.002       & $-$0.031 &$-$0.005 &   +0.032 & +0.098  &0.101    &0.099      \\
$[$Al/Fe$]${\sc i}&  2 &  +0.043       & $-$0.036 &$-$0.024 &   +0.008 & +0.120  &0.121    &0.130      \\
$[$Si/Fe$]${\sc i}&  5 &$-$0.046       &   +0.023 &$-$0.003 &   +0.031 & +0.076  &0.079    &0.093      \\
$[$Ca/Fe$]${\sc i}& 14 &  +0.063       & $-$0.070 &$-$0.015 & $-$0.012 & +0.045  &0.049    &0.088      \\
$[$Sc/Fe$]${\sc ii}& 6 &$-$0.008       &   +0.045 &  +0.008 & $-$0.007 & +0.069  &0.070    &0.071      \\
$[$Ti/Fe$]${\sc i}& 16 &  +0.089       & $-$0.042 &$-$0.010 & $-$0.036 & +0.043  &0.051    &0.111      \\
$[$Ti/Fe$]${\sc ii}& 5 &$-$0.021       &   +0.044 &  +0.008 & $-$0.001 & +0.076  &0.077    &0.081      \\
$[$V/Fe$]${\sc i} &  9 &  +0.092       & $-$0.044 &$-$0.004 & $-$0.021 & +0.057  &0.062    &0.120      \\
$[$Cr/Fe$]${\sc i}& 14 &  +0.057       & $-$0.041 &$-$0.015 & $-$0.001 & +0.045  &0.047    &0.080      \\
$[$Mn/Fe$]${\sc i}&  2 &  +0.037       & $-$0.039 &  +0.001 &   +0.022 & +0.120  &0.122    &0.128      \\
$[$Co/Fe$]${\sc i}&  4 &$-$0.005       &   +0.013 &$-$0.005 &   +0.002 & +0.085  &0.085    &0.085      \\
$[$Ni/Fe$]${\sc i}& 27 &$-$0.010       &   +0.017 &  +0.000 &   +0.004 & +0.033  &0.022    &0.035      \\
$[$Y/Fe$]${\sc i}  & 2 &  +0.104       & $-$0.025 &$-$0.019 &   +0.016 & +0.120  &0.121    &0.169      \\
$[$Zr/Fe$]${\sc i} & 6 &  +0.088       & $-$0.011 &$-$0.013 & $-$0.001 & +0.069  &0.070    &0.122      \\
$[$Zr/Fe$]${\sc ii}& 1 &  +0.079       & $-$0.034 &$-$0.006 &   +0.027 & +0.170  &0.171    &0.193      \\
$[$Ba/Fe$]${\sc ii}& 2 &  +0.017       &   +0.022 &  +0.015 & $-$0.045 & +0.120  &0.124    &0.123      \\
$[$La/Fe$]${\sc ii}& 3 &  +0.116       & $-$0.038 &$-$0.007 & $-$0.008 & +0.098  &0.100    &0.165      \\
$[$Ce/Fe$]${\sc ii}& 1 &  +0.107       & $-$0.044 &$-$0.007 &   +0.017 & +0.170  &0.171    &0.210      \\
$[$Eu/Fe$]${\sc ii}& 2 &  +0.095       & $-$0.040 &$-$0.007 &   +0.010 & +0.120  &0.121    &0.162      \\
\hline
\end{tabular}
\label{t:sensitivity}
\end{table*}

\section{Errors in the atmospheric parameters and cosmic scatter in Iron}

Following the tested procedure of previous papers, discussed in detail in
Paper V on the twin-cluster NGC~6441, similarly affected by problems of
differential reddening, we estimated 
individual (i.e. star-to-star) errors in the derived abundances 
by considering the three main error sources, i.e.
errors in temperatures, in microturbulent velocities and in the
measurements of $EW$s (the effects of errors in
surface gravities and in the adopted model metallicity are negligible on the
total error budget).

We started by evaluating the sensitivity of the derived abundances  on the
adopted atmospheric parameters. This was obtained by re-iterating the analysis
while varying each time only one of the parameters for the star 108895, assumed
as representative of our program stars; the results are listed in
Table~\ref{t:sensitivity}. In this Table, the first row shows the amount of the
variations in the parameters, the second and third rows list the estimated
star-to-star (internal) and systematic errors (that apply to the whole cluster)
in each parameters. 
The resulting sensitivities of abundances to variations in each parameter are
shown below, Columns from 3 to 6; in Column 2 we report the average (over all
the stars analyzed) number of lines used for each element.

The next step is to derive realistic estimates of internal and systematic 
errors in the atmospheric parameters and in measured $EW$s.

\paragraph{Internal errors}
In the following we will concentrate on the major error sources quoted above.

\paragraph{Errors in temperatures.} 
We adopted an internal error of 0.03 mag in $K$ magnitudes (from the mean
photometric uncertainties displayed on the 2MASS web site) and errors due to
the variable interstellar reddening across the cluster. According to Pritzl et
al. (2002) the rms associated to $E(B-V)$ is about 0.03 mag, leading to a
contribution of 0.012 mag on $K$, since A$_K \sim 0.4 E(B-V)$. 
In summary, the errors
in $K$ should not exceed 0.032 mag. When combined with the slope of the adopted
relation between T$_{\rm eff}$ and $K$ magnitude of about 192 K/mag, the
internal error in T$_{\rm eff}$ is estimated to be about 6~K.

\paragraph{Errors in microturbulent velocities.}
We repeated the analysis for star 101131 by changing $v_t$ until the 1$\sigma$
value\footnote{This value was derived as the quadratic mean of the 1 $\sigma$
errors in the slope of the relation between abundance and
expected line strength for all
stars.} from the original 
slope of the relation between line strengths and
abundances was reached; the corresponding internal error in microturbulent
velocities is 0.07 km~s$^{-1}$. 

\paragraph{Errors in measurement of equivalent widths.}
In order to estimate this contribution, we assumed that the average rms 
scatter (0.170 dex) in Fe {\sc i} abundance for our stars is the typical error
in abundance derived from an individual line. When divided by the square root
of the typical average  number of measured lines (89), this provides a typical
internal error of 0.018 dex as the contribution to the error due to
uncertainties in $EW$ measurement for individual lines. 

\paragraph{}
Once the internal errors are combined with sensitivities of 
Table~\ref{t:sensitivity}, the derived
individual star errors for Fe amount to 0.001 dex and 0.034 dex due
to the quoted uncertainties in T$_{\rm eff}$ and $v_t$.

Adding in quadrature the contribution due to errors in $EW$s 
we may evaluate the expected scatter in [Fe/H] due to the most relevant 
uncertainties in the analysis. We 
derive $\sigma_{\rm FeI}$(exp.)=$0.038 \pm 0.014$ dex (statistical error). The
inclusion of contributions due to uncertainties in surface gravity or model
metallicity (that results in the values reported in Table~\ref{t:sensitivity},
Column 8) does not alter our conclusions. 

Total internal and systematic errors are reported in Table~\ref{t:sensitivity} 
in Cols. 8 and 9 respectively, for iron and the other elements analyzed.
They are obtained by summing quadratically the contributions of the
individual sources of errors, scaled according to the proper estimates of the
real amount of errors in NGC~6388\footnote{For the systematic errors, the
contribution due to $EW$s and $v_t$ (quantities derived from our own analysis)
were weighted by dividing by the square root of number of stars observed in
NGC~6388}. We neglected the effects of covariances between the error sources;
these effects are expected to be very small for our program stars.

On the other hand, the observed star-to-star scatter in NGC~6388 is virtually
the same of the estimated internal error; we have
$\sigma_{\rm FeI}$(obs.)=$0.038 \pm 0.014$ (statistical error), obtained as
the rms scatter of abundances from Fe {\sc i} lines.
The conclusion from our dataset of high resolution spectra is that 
the observed star-to-star rms scatter in Fe abundances in NGC~6388 is no
more than 9\%, it is fully compatible with the major error sources entering in
the abundance analysis, suggesting to exclude any intrinsic metallicity spread
in this cluster. We will return on this issue with the analysis of the
GIRAFFE/FLAMES data (Carretta et al. 2006, in preparation).

\paragraph{Systematic errors}
They are not relevant within a single cluster, but could
be used when comparing different GCs, as in the present project.

Briefly, errors in the adopted reddening translate into systematic errors in
derived temperatures, since we adopted a
relation between temperatures and magnitude $K$, based on turn
on $T_{\rm eff}$ from dereddened $V-K$, through the calibrations by Alonso et
al. 
When combined with the slope $\sim 670$ K/mag of the $T_{\rm eff}$ vs $V-K$
relation, the adopted error of about 0.03 mag in $E(B-V)$ results into  a
systematic uncertainty of $\pm 55$~K for the cluster, increasing to $\pm
57$~K if we add in quadrature another contribution of  0.02 mag error in the
zero point of the $V-K$ color.

Errors in surface gravity might be obtained by propagating uncertainties in
distance modulus (about 0.1 mag), stellar mass (a conservative 10\%) and the
above error of 57 K in effective temperature. The quadratic sum results into a
0.062 dex of error in $\log g$. 
The systematic error relative to the microturbulent velocity $v_t$, a quantity
derived from our own analysis, was divided by the square root of the number of 
the observed stars, and is estimated in 0.026 km/s.

By using again sensitivities in Table~\ref{t:sensitivity}, these
contributions of systematic errors in $T_{\rm eff}$, $\log g$ and
$v_t$ can be translated in errors in the metallicity. 
Adding in quadrature the
statistical error of individual abundance determinations, 0.014 dex, we end up
with  a total systematic error of 0.025 dex.

Hence, we conclude that on the scale we are defining throughout this series  of
papers, the metal abundance of NGC~6388 is [Fe/H]$=-0.44\pm 0.014\pm 0.025$
dex, where the first error bar refers to the individual star-to-star errors and
the second one is relative to the cluster (or systematic) error.

\paragraph{A note about scale errors}
Scale errors are difficult to be estimated properly, and they are certainly 
much larger than the individual (internal) star and cluster (systematic) errors
mentioned  above. 
There are various
reasons for significant scale errors, including systematic measurement errors
(e.g. inappropriate location of the reference continuum level, or line
fitting), in the abundance analysis (damping parameters), uncertainties in the
adopted temperature scale (the Alonso et al. scale), in the modeling of the
stellar atmospheres (we are using 1-d Kurucz models, which neglect e.g.
horizontal inhomogeneities in stellar atmospheres and cooling of the outer
layers due to adiabatic expansion), and possible non-LTE effects in the
formation of Fe lines. All these errors are likely to be quite similar in all
stars considered in this series of paper, possibly with some trends with
luminosity, temperature or metal abundance. We think that a full estimate of
these uncertainties is not needed here, insofar we refer our metal abundances
to our own scale. However, when transforming our [Fe/H] values in absolute
abundances, as those e.g. required for modeling of stellar evolution, the
possibility of significant errors should be kept in mind.

We will examine this issue in more detail when data from the whole sample of
clusters will have been analyzed.

\section{Other elements}

Abundances for other elements measured in NGC~6388 are listed in
Table~\ref{t:protonu63} to Table~\ref{t:ncaptu63a}. For each element in
individual stars we give the number of lines used, the average abundance and
the rms scatter of individual values.
In Table~\ref{t:bulgegc} we give the average abundances for the cluster, as
well as the rms scatter of individual values around the mean value (columns 2
and 3). In columns 4 and 5 
of this Table we report the analogous values derived in Paper III
for the twin cluster NGC~6441 from high resolution UVES-FLAMES spectra. In the
other columns of this Table we list as a comparison the abundance ratios for
two other bulge clusters studied by our group (NGC~6528, Carretta et al. 2001;
NGC~6553, Cohen et al. 1999), as well as for the metal-rich cluster Ter 7 
considered part of the Sgr dwarf galaxy (Sbordone et al. 2005, Tautvaisiene et
al. 2004). Whenever possible (e.g. for NGC~6528, NGC~6553 and the analysis by
Sbordone et al.) element ratios have been recomputed by using the set of solar
reference abundances used in the present work (see Gratton et al. 2003).

\subsection{The proton-capture elements}

In Table~\ref{t:protonu63} we summarize the abundances for the elements
involved in proton-capture reactions at high temperature (see Gratton et al.
2004 for a review) in NGC~6388.

Abundances of O, Na, Mg and Al rest on measured $EW$s. 
The derived Na abundances were corrected for effects of
departures from the LTE assumption using the prescriptions by Gratton et al.
(1999).

\begin{table*}
\centering
\caption[]{Abundances of proton-capture elements in stars of NGC~6388; 
n is the number of lines used in the analysis.}
\begin{tabular}{rrccrccrccrcc}
\hline
Star        & n & [O/Fe] & rms & n & [Na/Fe] & rms & n & [Mg/Fe] & rms & n & [Al/Fe] & rms \\ 
            &   & (dex)  &     &   &  (dex)  &     &   & (dex)   &     &   &  (dex)  &     \\   
\hline
 77599 & 2 &$-$0.33 &0.14 & 4 & +0.71	&0.25 & 2 &  +0.16  &0.02 & 2 &  +0.98  &0.12 \\ 
 83168 & 2 &$-$0.20 &0.16 & 4 & +0.49	&0.18 & 3 &  +0.16  &0.08 & 2 &  +0.74  &0.02 \\ 
101131 & 2 &$-$0.58 &0.05 & 4 & +0.76	&0.17 & 3 &  +0.18  &0.01 & 2 &  +0.87  &0.24 \\ 
108176 & 2 &$-$0.26 &0.14 & 3 & +0.72	&0.12 & 3 &  +0.34  &0.04 & 2 &  +0.55  &0.15 \\ 
108895 & 2 &$-$0.42 &0.05 & 4 & +0.50	&0.16 & 3 &  +0.19  &0.05 & 2 &  +0.68  &0.14 \\ 
110677 & 2 &$-$0.12 &0.06 & 3 & +0.34	&0.09 & 2 &  +0.18  &0.03 & 2 &  +0.23  &0.01 \\ 
111408 & 2 &$-$0.17 &0.16 & 4 & +0.66	&0.16 & 2 &  +0.24  &0.06 & 2 &  +0.77  &0.12 \\ 
\hline
\end{tabular}
\label{t:protonu63}
\end{table*}

Oxygen abundances are obtained from the forbidden [O {\sc i}] lines at 6300.3
and 6363.8~\AA; the former was cleaned from telluric contamination by
H$_2$O and O$_2$ lines using a synthetic template (see Paper I for more
details).
Neither
CO formation nor the high excitation Ni {\sc i} line at 6300.34~\AA\
are a source of concern: as discussed in Paper III, the corrections due to
neglecting CO formation in these stars and to the (small) contribution of the
blending Ni {\sc i} line are of the order of about 0.05 dex and they
roughly compensate with each other, being of opposite signs.

From Table~\ref{t:bulgegc} we note that the rms scatter of the average
abundance is generally in good agreement with those estimated by the error
analysis in Sect. 4, apart from cases where the observed scatter is large due
to the small number of available lines. 
However, the observed scatter for proton-capture elements 
is clearly exceeding the predicted observational errors, apart from the case of
Mg, indicating that an intrinsic scatter is present in these abundance ratios.

\begin{figure}
\centering
\includegraphics[bb=20 140 225 716, clip, scale=0.80]{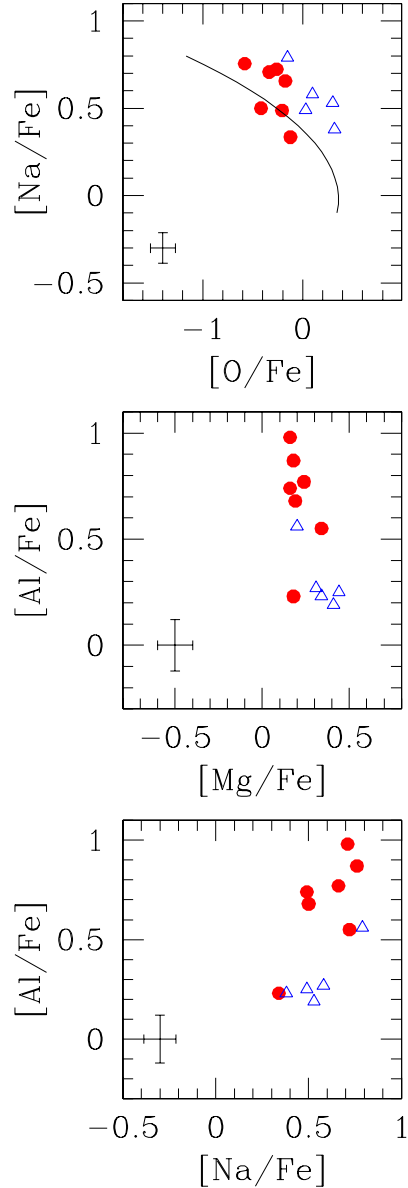}
\caption{Upper panel: [Na/Fe] ratios as a function of [O/Fe] for red giant 
stars in NGC~6388 from the present study (filled circles) and in NGC~6441
from Paper III (empty triangles). Middle panel: [Al/Fe] ratios as a function of
[Mg/Fe] ratios; symbols have the same meaning. Lower panel: [Al/Fe] ratios as a
function of [Na/Fe] ratios; symbols are as in previous panels.  
The random error bars take into account the uncertainties in
atmospheric parameters and $EW$s.}
\label{f:protonu63}
\end{figure}

The [Na/Fe] ratio as a function of [O/Fe] ratio is displayed in the upper panel
of Figure~\ref{f:protonu63} for the seven stars observed with UVES in NGC~6388
(filled circles) and, as a comparison, for the five stars in NGC~6441 from
similar data (Paper III, empty triangles).
The overall random error in O and Na as due to the contribution of errors in
the adopted atmospheric parameters and measurements of $EW$s in the present
paper is also shown.

The curve in Figure~\ref{f:protonu63} (from Fig. 5 of Carretta et al. 2006a) 
represents the typical pattern of the Na-O anticorrelation in globular clusters,
as defined on NGC~2808 and found generally appropriate for GCs
from a collection of about 400 stars in almost 20 GCs. Although the
sample based on high-resolution UVES spectra is limited, we may conclude that
stars in NGC~6388 participate to the classical Na-O anticorrelation, 
as already found also for giants in the twin cluster NGC~6441. The Na-O
anticorrelation will be studied in details separately, from a more extended 
sample based on Giraffe spectra in NGC~6388.

In the middle and lower panels of Figure~\ref{f:protonu63} we show the Al-Mg
anticorrelation and the Na-Al correlation in NGC~6388, as compared to the ones
in NGC~6441 (symbols are the same as in previous panels). Both these features
are predicted to set up as a consequence of H-burning at high temperatures
involving the NeNa and MgAl cycles.
The small spread observed in [Mg/Fe] is not surprising: as discussed in
Carretta et al. (2004), there is evidence suggesting that the MgAl cycle could
be less efficient in metal-rich clusters like 47~Tuc and M~71. Our findings for
NGC~6388 and NGC~6441 seem to support this view.

On the other hand, the spread observed in Al is quite robustly established, 
and it is a clearcut evidence of the presence (although limited) of products
of proton-captures in the MgAl cycle. The fraction of
Al-rich and Al-poor stars seems to be reversed in NGC~6388 and NGC~6441: the
former has mostly stars in the first group, whereas the majority of stars in
NGC~6441 show only a moderate Al-enhancement. 
Admittedly, the statistics is small, but this sort of dichotomy
in the Na-Al correlation was already noticed by Carretta (2006; his Fig. 4), 
using more extended samples of RGB stars in several GCs.

\subsection{$\alpha-$elements}

Abundances of elements from $\alpha-$processes in stars of NGC~6388 are listed
in Table~\ref{t:alphau63}.

\begin{table*}
\centering
\caption[]{Abundances of $\alpha$-capture elements in stars of NGC~6388. 
n is the number of lines used in the analysis.}
\begin{tabular}{rrccrccrccrcc}
\hline
Star        & n &[Si/Fe]{\sc i}& rms & n & [Ca/Fe]{\sc i} & rms & n & [Ti/Fe]{\sc i} & rms & n & [Ti/Fe]{\sc ii} & rms \\ 
            &   & (dex)  &     &   &  (dex)  &     &   & (dex)   &     &   &  (dex)  &     \\   
\hline
 77599 & 4 &  +0.34 &0.12 & 14 &  +0.14 &0.26 & 15& +0.38 &0.16 & 6 & +0.46  &0.31 \\ 
 83168 & 7 &  +0.37 &0.21 & 14 &$-$0.05 &0.19 & 15& +0.24 &0.17 & 4 & +0.19  &0.27 \\ 
101131 & 6 &  +0.42 &0.14 & 13 &  +0.10 &0.16 & 18& +0.41 &0.15 & 5 & +0.18  &0.25 \\ 
108176 & 4 &  +0.21 &0.15 & 13 &  +0.10 &0.20 & 15& +0.54 &0.14 & 5 & +0.22  &0.16 \\ 
108895 & 6 &  +0.25 &0.15 & 18 &  +0.04 &0.19 & 17& +0.34 &0.18 & 6 & +0.26  &0.28 \\ 
110677 & 4 &  +0.46 &0.09 & 13 &  +0.04 &0.20 & 16& +0.27 &0.17 & 5 & +0.36  &0.25 \\ 
111408 & 6 &  +0.21 &0.18 & 15 &  +0.08 &0.18 & 19& +0.39 &0.17 & 5 & +0.43  &0.19 \\ 
\hline
\end{tabular}
\label{t:alphau63}
\end{table*}

Apart from Ca, the elements ratios of the $\alpha-$elements 
(Mg, Si, Ti {\sc i} and {\sc ii}) show a clear overabundance with respect to
the solar values. For Ca we obtain low abundances, as in Paper III for NGC~6441;
however, as discussed in that paper, this value might not be representative
of the real Ca abundance; it might be an artifact of the
analysis, due to the rather cool temperature of stars analyzed (see Paper III).

The difference for abundances of Ti derived from neutral and singly ionized
lines (the latter not available in NGC~6441 where we restricted the analysis to
lines measured in the redder part of the UVES spectra) allows us to further
check the reliability of the adopted parameters. On average, we found that
[Ti/Fe]{\sc ii}$-$[Ti/Fe]{\sc i}$=-0.07$ with $\sigma=0.16$ dex. This slight
difference is not significant, supporting the goodness of the analysis.

An extensive set of abundances of $\alpha-$elements obtained with an
homogeneous and self consistent procedure permits to check the scenario where
dense objects may retain the ejecta of core collapse supernovae and maintain
an independent chemical evolution resulting in self-enrichment. 
At present we have available a comparison set made of i) field stars analyzed 
by Gratton et al. (2003b) with the same line parameters and solar reference
abundances used in the present series; ii) the field, presumably bulge, stars
analyzed in Paper III; iii) a thick disk globular cluster (47
Tuc, [Fe/H]$=-0.67$ dex), studied with the same approach by Carretta et al.
(2004); iv) two moderately metal-rich bulge clusters (NGC~6388 and NGC~6441,
[Fe/H]$=-0.44$ and $-0.43$ dex, respectively from the present work and Paper
III); v) and finally two bulge clusters at metallicity near or above solar
(NGC~6528, [Fe/H]=$+0.07$ dex, from Carretta et al. 2001, and NGC~6553,
[Fe/H]$=-0.16$ dex, from Cohen et al. 1999, as revised in Carretta et al.
2001). For the two latter clusters, the  $\alpha-$element ratios of the
original papers were corrected 
here for the different solar reference abundances adopted.

The metallicity range spanned by these clusters is the best suited for this
test, since studies of field stars show that
a significant contribution by type Ia SNe  is
noticeable  already at [Fe/H]$\sim -1$ dex, lowering the
[$\alpha$/Fe] ratio. We could then expect that for metallicities larger
than this limit, GCs
present an higher $\alpha$-element content if some degree of self-enrichment
does exist (see also discussion and references in Carretta et al. 2004).

\begin{figure}
\centering
\includegraphics[bb=20 140 590 700, clip, scale=0.43]{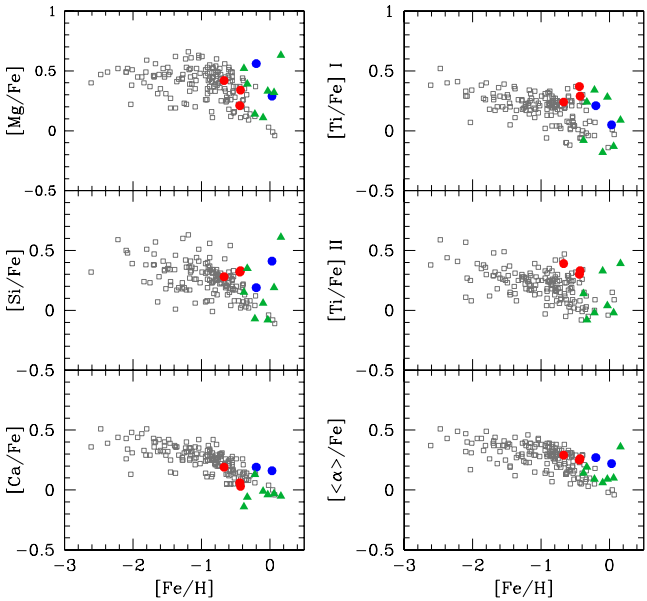}
\caption{Overabundance of $\alpha-$elements and of the average Mg+Si+Ca+Ti as
a function of the metallicity [Fe/H]. Open squares are field stars from Gratton
et al. (2003b); green triangles are field (bulge) stars from
Gratton et al. (2006). Big filled circles are metal-rich clusters, located according
to their metallicity: [Fe/H]$=-0.67, -0.44, -0.43, -0.16, +0.07$ dex for 47 Tuc,
NGC~6388, NGC~6441 (in red), NGC~6553 and NGC~6528 (in blue), respectively.}
\label{f:field3}
\end{figure}

A hint that this is what's really happening seems to come from 
Figure~\ref{f:field3}, where we superimposed the average abundance ratios of all
analyzed $\alpha$-elements in these GCs to field stars. The average ratio
[(Mg+Si+Ca+Ti)/Fe] is also shown in the bottom-right panel. More observations
are needed to firmly decide whether a real difference does exist; however, the
impression is that for several elements (Si, Ti {\sc i} and {\sc ii} and for 
the average $<\alpha>$) the position of the GCs is near the upper envelope of
the field distribution. Moreover, the differences between clusters and field
stars seem to increase as the metallicity increases (see the
behavior of NGC~6553 and NGC~6528).
This is what we expect if
the efficiency of clusters in retaining ejecta of core-collapse SNe is
approximatively constant, whereas the increased contribution of type Ia SNe is
decreasing more and more the [$\alpha$/Fe] ratio in the less dense field
environment.

\subsection{Fe-group elements}

Abundances of elements of the Fe-group are reported in Tables~\ref{t:fegu63}
and ~\ref{t:heavyu63}. We measured lines of Sc {\sc ii}, V {\sc i}, 
Cr {\sc i}, Mn {\sc i}, Co {\sc i} and Ni {\sc i}. Corrections due to the
hyperfine structure were applied in the relevant cases (Sc, V, Mn, Co; see
Gratton et al. 2003a for details). 

\begin{table*}
\centering
\caption[]{Abundances of Fe-group elements in stars of NGC~6388.} 
\begin{tabular}{rrccrccrccrcc}
\hline
Star        & n &[Sc/Fe]{\sc ii}& rms & n & [V/Fe]{\sc i} & rms & n & [Cr/Fe]{\sc i} & rms & n & [Mn/Fe]{\sc i}& rms \\ 
            &   & (dex)  &     &   &  (dex)  &     &   & (dex)   &     &   &  (dex)  &     \\   
\hline
 77599 & 3&   +0.06 &0.22 & 8 &+0.24 &0.21 & 12&  +0.04& 0.30 & 4&$-$0.23 &0.17 \\ 
 83168 & 6&   +0.13 &0.21 & 9 &+0.32 &0.17 & 18&$-$0.05& 0.21 & 4&$-$0.23 &0.19 \\ 
101131 & 6&   +0.04 &0.17 & 9 &+0.53 &0.17 & 10&$-$0.14& 0.27 & 3&$-$0.24 &0.25 \\ 
108176 & 7&   +0.05 &0.25 & 8 &+0.50 &0.17 & 11&  +0.14& 0.24 & 3&$-$0.26 &0.01 \\ 
108895 & 7&   +0.03 &0.19 & 9 &+0.38 &0.25 & 18&  +0.02& 0.26 & 3&$-$0.30 &0.14 \\ 
110677 & 7&   +0.11 &0.21 & 8 &+0.35 &0.25 & 19&$-$0.19& 0.22 & 4&$-$0.25 &0.19 \\ 
111408 & 4& $-$0.07 &0.15 &10 &+0.42 &0.19 & 14&$-$0.05& 0.22 & 3&$-$0.23 &0.20 \\ 
\hline
\end{tabular}
\label{t:fegu63}
\end{table*}

\begin{table}
\centering
\caption[]{Abundances of the Fe-group elements Co and Ni in stars of NGC~6388.}
\begin{tabular}{rrccrccrcc}
\hline
Star        & n &[Co/Fe]{\sc i}& rms & n & [Ni/Fe]{\sc i} & rms \\ 
            &   & (dex)        &     &   &  (dex)         &     \\   
\hline
 77599 & 4&   +0.15 &0.24 &24 &  +0.03 &0.22  \\ 
 83168 & 4&   +0.03 &0.17 &30 &  +0.03 &0.22  \\ 
101131 & 4&   +0.07 &0.16 &27 &  +0.01 &0.16  \\ 
108176 & 5&   +0.02 &0.25 &22 &  +0.03 &0.23  \\ 
108895 & 4& $-$0.07 &0.14 &30 &$-$0.02 &0.15  \\ 
110677 & 4& $-$0.01 &0.96 &29 &  +0.08 &0.21  \\ 
111408 & 4&   +0.10 &0.14 &28 &  +0.08 &0.21  \\ 
\hline
\end{tabular}
\label{t:heavyu63}
\end{table}

In general, Fe-group elements are clustered around the solar abundance ratio
[El/Fe]$\sim0.0$ dex. The overall pattern of these elements seems to be shared
by all bulge clusters analyzed by us (see Table~\ref{t:bulgegc}) and also by 
the globular cluster Terzan 7, likely to belong to the Sgr dwarf galaxy and
used in Paper III as a comparison. Abundance
ratios for the iron group elements in this cluster are taken from 
Tautvaisiene et al. (2004) and they agree with the solar ratios.

Special attention is deserved by Mn, the only Fe-peak element (with copper)
clearly showing an abundance deficiency. The nucleosynthetic situation for this
odd Fe-group element is still unclear; the most commonly accepted hypothesis to
explain the observed pattern of [Mn/Fe] vs [Fe/H] (see Carretta et al. 2004 
and Sobeck et al. 2006 for a summary of recent data and detailed discussions)
can be briefly summarized as follows: (i) the fraction of Mn
contributed to the interstellar medium by SNe Ia is larger than the fraction
produced by SNe II (Gratton 1989) or (ii) the Mn yields from SNe II are 
metallicity-dependent (Arnett 1971). Without entering this long standing
debate, we may try to locate the positions of the two peculiar bulge cluster
NGC~6388 and NGC~6441 by using a limited but very homogeneous set of data.

\begin{figure}
\centering
\includegraphics[bb=30 150 570 700, clip, scale=0.45]{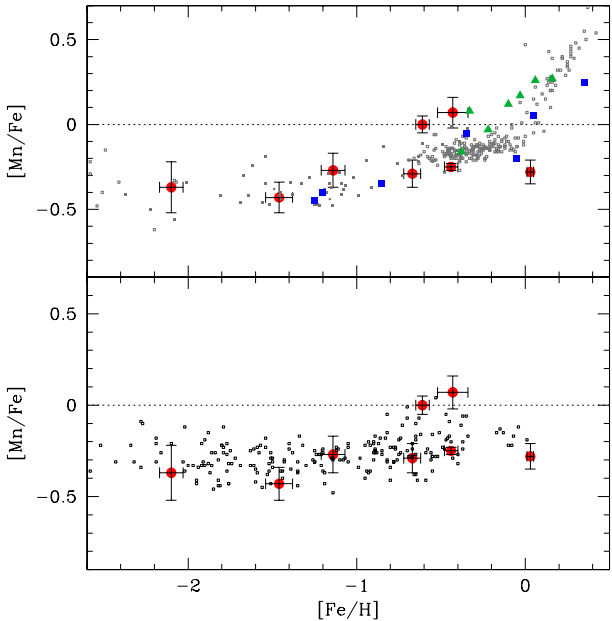}
\caption{Upper panel: [Mn/Fe] ratio as a function of the metallicity [Fe/H].
Large filled circles are GCs analyzed homogeneously by our group plus Ter 7
(Tautvaisiene et al. 2004, at  [Fe/H]=$-0.61$).  In order of increasing  [Fe/H]
values our GCs are: NGC~6397, NGC~6752, NGC~2808, 47~Tuc, NGC~6388, NGC~6441
and NGC~6528; references are given in the text.  
Error bars for GCs represent the $rms$ scatter about the average.
The field stars used as reference are from the following sources: 
small filled squares, Gratton et al. (2003b); 
small open squares, extremely metal-poor stars ([Fe/H]$\le -2$) from 
McWilliam et al. (1995) and Ryan et al. (1996), and metal-rich 
([Fe/H]$\ge ~-0.5$) disk stars from Feltzing \&
Gustafsson (1996) and Reddy et al. (2003); large filled (blue) squares, 
bulge stars from McWilliam et al. (2003); filled (green) triangles, bulge stars
from Gratton et al. (2006, Paper III).
Lower panel: the same GCs, but compared to the homogeneous re-analysis of field
stars by Sobeck et al. (2006). } 
\label{f:mnfe}
\end{figure}

In Figure~\ref{f:mnfe} (upper panel) we plot abundance ratios [Mn/Fe] for field
stars from Gratton et al. (2003b); to extend the comparison into the metal-poor
and metal rich domains, we employed studies of McWilliam et al. (1995), Ryan et
al. (1996), Feltzing and Gustafsson (1998) and Reddy et al. (2003).  As a
reference sample of bulge field stars we added the stars analyzed by 
McWilliam et al. (2003) and by Gratton et al. in Paper III. We made no attempt
to normalize all the results to a common scale; however, there are no 
pronounced offsets between literature sources and the sample  by Gratton et al.
(2003b), analyzed with procedures very similar to the one  adopted in the
present series. This makes us confident about the self-consistency of the
dataset of field comparison stars.

We restrict the sample of GCs to those spanning the whole range of metallicity
typical of Galactic globular clusters, but all studied by our group with an
extremely homogeneous analysis (procedures, packages, atomic parameters). In
Figure~\ref{f:mnfe} (upper panel) we plot the average values 
(in order of increasing [Fe/H]) for the globular cluster: NGC~6397 and 
NGC~6752 (Carretta \& Gratton 2006, in prep.); NGC~2808 (Carretta 2006); 47~Tuc
(Carretta et al. 2004); NGC~6388 (present work); NGC~6441 (Paper III) and
NGC~6528 (Carretta et al. 2001). Again, we corrected the values for NGC~6528 to
take into account the different solar abundances used in that work. Finally, as
a representative of a likely accreted component, we added the values for 
Terzan~7 ([Fe/H]$=-0.61$, [Mn/Fe]=0.0) from Tautvaisiene et al. (2004).

From this comparison we can see that metal-poor clusters ([Fe/H]$\leq -0.7$)
are in very good agreement with the more or less constant plateau at
[Mn/Fe]$\sim -0.3\div -0.4$ dex as defined by field stars, including the
metal-poor bulge component studied by McWilliam et al. (2003).
However, for larger metallicities, there is a hint that Mn
abundances in cluster and field stars may not vary in lockstep. In particular,
NGC~6388 and NGC~6528 show an underabundance of Mn increasingly deviating 
with respect to field stars. On the other hand, in NGC~6441 the [Mn/Fe] ratio 
is about 0.2 dex above the mean [Mn/Fe] value of field stars of similar 
metallicity.
We note that NGC~6388 and NGC~6528 seem to continue the plateau at constant
[Mn/Fe] up to solar metal abundance. The pattern shown by NGC~6441 is more 
similar to that found in the Sagittarius cluster Ter~7; it
is however not clear if this could indicate also a similar (independent) 
chemical evolution,
since the average value found by Tautvaisiene et al. (2004) in Ter~7 for the
$\alpha$-elements is [$\alpha$/Fe]$=+0.08\pm 0.04$ dex, about 0.2 dex below the
mean value found in NGC~6441 (and NGC~6388). Moreover, some offsets might
exist, because we do not know what solar reference abundances were used for
Ter~7.

To limit the importance of possible offsets among different data sets,  
in Figure~\ref{f:mnfe} (lower panel) we plot the same sample of GCs, but 
compared to the most recent, complete and homogeneous re-analysis of Mn
abundances in field stars (mostly in the metallicity range 
$-2.5 \le$[Fe/H]$\le -0.7$) by Sobeck et al. (2006). We applied to their data
a small offset (+0.07 dex) to take into account differences in solar reference
abundances.
In this analysis the pattern of underabundant [Mn/Fe] ratios extends up
to solar metallicity, and all GCs have on average very similar Mn abundances,
except for Ter~7 and NGC~6441. If we exclude these two objects, globular
clusters have a mean value of [Mn/Fe]$=-0.31$ dex (rms=0.07 dex, 6 clusters)
which is virtually the same within the uncertainties of the average value of
field stars ([Mn/Fe]$=-0.29$ dex, rms=0.08 dex, when using our adopted solar
abundances). This results strongly supports the finding of Sobeck et al. (2006)
about the invariance of the relative contribution to the two populations from
core collapse supernovae.

On the other hand, both NGC~6441 and Ter~7 stand well off from this average
Mn value, by more than 5$\sigma$ the first and more than 4$\sigma$ the second.
As discussed in Paper III, the almost solar ratio Mn/Fe found in NGC~6441 might
indicate a different chemical history, maybe in independent system lately
accreted in the Galaxy. However, things are not so simple, because also
Ter~7, a cluster supposedly attributed to the Sgr dwarf galaxy, shows a solar
[Mn/Fe] ratio. This last result does contrast with the Mn content found by
McWilliam et al. (2003) to be underabundant in metal-rich stars of the same Sgr
dwarf spheroidal galaxy.

Further insight in this metallicity regime is required to clarify the run and
the nucleosynthetic history of Mn, together with informations to accurately
disentangle the mix of stellar populations (thin disk, bulge and likely
accreted stars) in this range of [Fe/H].

\subsection{Neutron-capture elements}

Finally, the abundances of elements produced by neutron-capture reactions are
listed in Table~\ref{t:ncaptu63} and Table~\ref{t:ncaptu63a}. In stars of
NGC~6388 we measured: the light $s-$process elements Y {\sc i}, Zr {\sc i} 
and {\sc ii} of the first neutron peak, the heavier $s-$process species
Ba {\sc ii}, La {\sc ii} and Ce {\sc ii}, belonging to the second neutron peak,
and the element Eu {\sc ii}, the classical
surrogate for $r-$process nucleosynthesis. 

The importance of $\alpha$-elements stay in understanding the relative weight of the
nucleosynthesis contributions from core-collapse SNe and lower mass SNe Ia in
binary systems with longer lifetimes. 
This is mirrored by the relevance of the abundance ratios of
n-capture elements to disentangle the contributions and possible metallicity
dependences in yields from massive stars with respect to intermediate mass
stars. In fact, $r-$process elements are predominantly produced in the last
phases of the life of high mass stars (together with the $weak$ componente of
$s-$elements), whereas the quiescent He burning in low and intermediate-mass
stars is chiefly responsible for producing elements from the $main$ $s-$process
component.

In particular, the analysis of n-capture elements can offers key observational
constraints about model of formation and early evolution of GCs. James et al.
(2004) performed the most extensive study of n-capture elements in unevolved
cluster stars, coupling their results to data collected from literature, to
discuss the possible scenario for self-enrichment models in GCs. However, their
data reached only [Fe/H]$\sim -0.7$ (47 Tuc) and more metal-rich clusters are
needed to explore the classical self-enrichment scenario (e.g. Cayrel 1986),
following the suggestion by Truran (1988). As already discussed in the Sect.
5.2, at nearly solar metallicities should be enhanced the contrast between two
possible, competing patterns of abundance: (i) the one imprinted by 
self-enrichment phenomena in old objects like the GCs, reflecting mainly the
nucleosynthetic contribution from high mass stars only, and (ii) the cumulative
effects of chemical enrichment from several stellar generations, increasingly
dominated by yields from long lived, less massive progenitors, whose net effect 
is to lower the pristine abundance ratios typical of low metal deficient stars.

The analysis of NGC~6388 is particularly important, as it is one of the 
most metal-rich globular clusters known in our Galaxy. Thus it can be a very 
powerful complement to the work of James et al. (2004) to test 
the self-enrichment scenario, as suggested by Truran (1988).

In typical self-enrichment scenarii (Cayrel 1986; Truran 1991), but also in 
more recent developments (e.g. Thoul et al. 2002, and other improvements in the
Evaporation/Accretion/Self-Enrichment EASE scenario), globular cluster stars are formed in a way that is similar to
very old and metal-poor halo field stars, and should consequently show a
clear $r$--process dominated origin for their heavy elements (e.g. [Ba/Eu]
ratios close to a pure $r$--process only value, see discussion in James et al.
2004). This should be true also for the most metal-rich globular clusters. 
Among the best suited elements to test these scenarios are those formed by 
neutron-capture reactions.

\begin{figure}
\centering
\includegraphics[bb=30 150 570 700, clip, scale=0.45]{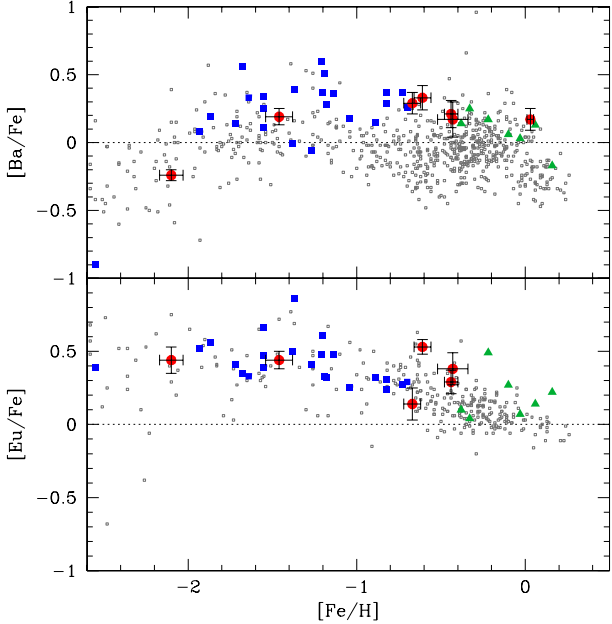}
\caption{Upper panel: [Ba/Fe] ratio as a function of the metallicity [Fe/H].
Large filled circles (in order of increasing metallicity) are GCs analyzed by 
James et al. (2004; NGC~6397, NGC~6752 and 47 Tuc), 
Tautvaisiene et al. (2004; Ter~7 at [Fe/H]=$-0.61$), this work (NGC~6388),
Gratton et al. (2006, Paper III; NGC~6441) and Carretta et al. (2001;
NGC~6528). Error bars for GCs represent the $rms$ scatter about the average.
The field stars used as reference are from the following sources: Venn et al.
(2004; small empty squares), James et al. (2004; large filled squares) and
Gratton et al. (2006; large filled triangles).
Lower panel: the same but for the [Eu/Fe] ratio as a function of [Fe/H]. In
this panel NGC~6528, with no Eu measurements, is missing.}
\label{f:baeufe}
\end{figure}

In Figure~\ref{f:baeufe} we show as a function of metallicity the run of Ba 
and Eu abundances as typical elements produced by $s-$ and $r-$process.
We restricted again the sample of GCs (large filled circles) mainly to those
analyzed by us (plus Ter~7 from Tautvaisiene et al. 2004 as an example of
accreted object). As reference field stars we plot data from the extensive
compilation by Venn et al. (2004; small empty squares), from the
self-consistent analysis of field stars by James et al. (2004; not included in
the compilation by Venn et al.; filled squares)
and from Paper III (putative bulge field stars, filled triangles).
For clusters analyzed in James et al. (2004) we adopted the weighted mean of
values obtained from subgiants and turn-off stars for all elements.

Eu is a probe of almost pure $r-$process nucleosynthesis at any metal-abundance
and the lower panel of Figure~\ref{f:baeufe} strongly supports the common
origin of elements from $\alpha$ and rapid neutron capture in the same stellar
environment (the death of high mass stars). As for [$\alpha$/Fe] ratios, the 
resulting pattern of [Eu/Fe] is a flat plateau dominated by $r-$process
contribution at low metallicity ([Fe/H]$<-1$) with a well defined knee when 
the contribution of Fe from type Ia SNe becomes increasingly overwhelming.

Two features must be noted in this diagram. The first is that for GCs the
[Eu/Fe] values show a quite small dispersion, the average ratio being
[Eu/Fe]=+0.37 with $\sigma=0.14$ dex. The value for 47 Tuc seems a bit lower:
excluding this cluster we would obtain [Eu/Fe]=+0.42 with $\sigma=0.09$ dex. 
There seems to be no selection bias involved in this result: from a larger
sample of 20 clusters in literature Gratton et al. (2004) obtained a mean value
of +0.40 dex ($\sigma=0.13$). This results extends the findings by James et al.
(2004) up to the high metallicity of NGC~6388 and NGC~6441: the [Eu/Fe] ratios 
are very similar for GCs of any metal abundance. The bottom line is that these
Eu abundances were already established in the interstellar medium (ISM) at the 
epoch when most globular clusters formed. 
The second point to notice is that GCs follow rather well the pattern of field
stars, but only up to the metallicity of 47 Tuc. Apparently, for higher [Fe/H]
values the [Eu/Fe] ratios still run flat for GCs, whereas the ratio in field
stars is progressively decreased by the cumulative effect of injection in the
ISM of yields from type Ia SNe. This is not due to a bias in stellar population
(disk versus bulge): in this plot both NGC~6388 and NGC~6441 have a ratio 
about 0.2 dex higher than the average value of the putative bulge field stars
analyzed in Paper III. As discussed in that work, this could indicate a larger
contribution by core collapse SNe. Another explanation may be a higher
efficiency to retain ejecta of massive stars in strongly bound GCs: this is
supported by the fact that the more loose galactic open clusters seem to follow
rather well the pattern of [Eu/Fe] versus [Fe/H] defined by field stars (see 
Fig. 6 in Gratton et al. 2004). 

The larger dispersion of the [Ba/Fe] ratio as a function of the metallicity
(upper panel of Figure~\ref{f:baeufe}) reflects the more complex
nucleosynthetic history of Ba. As remarked in James et al. (2004), in general 
the [Ba/Fe] ratio in GCs seems to follow the progressive enrichment signature 
found in halo and disk field stars. This led to the conclusion that GCs have
been uniformely enriched in neutron capture elements not only by the
$r-$process, but also by $s-$process nucleosynthesis. 

\begin{figure}
\centering
\includegraphics[bb=30 150 570 700, clip, scale=0.45]{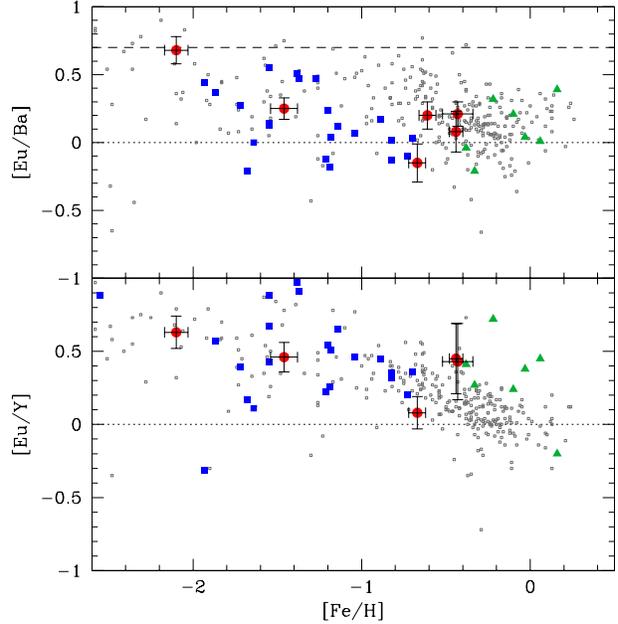}
\caption{Relative strengths of the $r-$process and $s-$process as a function of
metal abundance in Galactic globular clusters and field stars.  In the upper
panel the pure $r-$process element Eu is plotted relatively to Ba  (81\% from
$s-$process in the Sun, Arlandini et al. 1999), sampling the second  neutron
peak. The dotted line mark the total $r+s$ solar system ratio; the dashed line
represents the solar $r-$process only ratio [Eu/Ba]$_{(r-only)}=+0.70$
(Arlandini et al. 1999). In the lower panel Y (belonging to the first peak) is
used as $s-$process reference element (74\% in the Sun, Travaglio et al. 2004).
Data and symbols for GCs and  field stars are as in the previous Figure, apart
from NGC~6528, where Eu  was not measured, and Ter~7, with no measurements for
Y.}
\label{f:eubayfe}
\end{figure}

In Figure~\ref{f:eubayfe} the relative weight of elements from $r-$ and 
$s-$processes is examined for our
subsample of GCs and the reference field stars (the same as in
Figure~\ref{f:baeufe}). This is done by using the abundance ratios of Eu (a
pure $r-$process element) relatively to Y and Ba surrogates for $s-$process
elements in the first and second neutron peak.
At low metallicities the dominance from $r-$process nucleosynthesis is evident
in metal-poor stars as well as in clusters. Low metallicity GCs like NGC~6397
(James et al. 2004), but also M~15 and M~92 (Sneden et al. 19997, 2000) have
abundances of neutron-capture elements well matched by a scaled solar system
pure $r-$process abundance distribution ([Eu/Ba]$_{(r-only)}=+0.70$, Arlandini
et al. 1999). As the metallicity increases, field and GCs show ratios [$r$/$s$]
in rough agreement, although the scatter in [Eu/Ba] is much larger than for
[Eu/Fe] or even the similar ratio [Eu/Y]. 

Both panels in Figure~\ref{f:eubayfe}
show the typical run of decreasing ratios due to progressive enrichment of
elements (such as Ba and Y) forged within intermediate mass stars with longer
evolutionary timescales for the release of yields: a roughly flat plateau
followed by a decrease approaching solar metallicity (see also analogous
Figures in Pritzl et al. 2005). This is more evident in the lower panel, where
the scatter is reduced; the larger dispersion in the [Eu/Ba] ratios may be
due to different assumptions in the treatment of the hyperfine and isotopic
splitting and/or to the corrections for departures from LTE adopted in
different studies.

The [Eu/Ba] ratios in the metal-rich bulge clusters NGC~6388 and NGC~6441 are 
the same within the associated uncertainties in the analysis (see 
Table~\ref{t:bulgegc}) and follow the trend defined by field stars with the 
same metallicity. On the other hand, [Eu/Y] ratios in these GCs seem to be a
factor 2 larger than the average value for disk field stars, although still
compatible within the (rather large) associated error bars. 
These ratios are very similar to those of field and GCs at low metallicities, 
which would suggest a $r-$process dominated nucleosynthesis, with lack of
strong contributions from $s-$elements produced in AGB stars.
In comparison, also
the field bulge stars (Paper III) show a higher [Eu/Y] ratio, similar to the
one derived in NGC~6388 and NGC~6441. It is interesting to note that light
$s-$elements like Y do not follow always the pattern of the heavier neutron
capture elements (e.g. Burris et al. 2000); indeed they may also be produced by
an additional source ($weak$ $s-$process component) in the He-core burning
phase of massive stars (see Travaglio et al. 2004). Again, this would be
evidence of a larger contribution from massive stars nucleosynthesis in bulge
stellar populations. 

The abundance distribution of neutron capture abundance ratios in these two
metal-rich clusters seems to be far away from the pure $r-$process ratio and 
this evidence casts severe limitations to a scenario of primordial pure
self-enrichment only. The scenario for the formation and early evolution of
globular clusters is more complicate. 

\begin{table*}
\centering
\caption[]{Abundances of neutron-capture elements in stars of NGC~6388.} 
\begin{tabular}{rrccrccrccrcc}
\hline
Star        & n &[Y/Fe]{\sc i}& rms & n & [Zr/Fe]{\sc i} & rms & n & [Zr/Fe]{\sc ii} & rms & n & [Ba/Fe]{\sc i}& rms \\ 
            &   & (dex)  &     &   &  (dex)  &     &   & (dex)   &     &   &  (dex)  &     \\   
\hline
 77599 & 2& $-$0.15 &0.20 & 6 &$-$0.12 &0.22 & 1&$-$0.18&  & 3&  +0.13 &0.07 \\ 
 83168 & 2& $-$0.27 &0.12 & 6 &$-$0.21 &0.22 & 1&$-$0.03&  & 3&  +0.19 &0.17 \\ 
101131 & 2& $-$0.07 &0.03 & 6 &$-$0.15 &0.28 & 1&$-$0.05&  & 3&  +0.42 &0.15 \\ 
108176 & 2&   +0.10 &0.13 & 6 &$-$0.02 &0.33 & 1&$-$0.23&  & 3&  +0.15 &0.14 \\ 
108895 & 2& $-$0.40 &0.10 & 6 &$-$0.35 &0.17 & 1&$-$0.50&  & 3&  +0.10 &0.10 \\ 
110677 & 2& $-$0.34 &0.10 & 6 &$-$0.27 &0.27 & 1&$-$0.07&  & 3&  +0.18 &0.11 \\ 
111408 & 2&   +0.01 &0.20 & 6 &$-$0.04 &0.26 & 1&$-$0.33&  & 3&  +0.29 &0.13 \\ 
\hline
\end{tabular}
\label{t:ncaptu63}
\end{table*}

\begin{table*}
\centering
\caption[]{Abundances of other neutron-capture elements in stars of NGC~6388.} 
\begin{tabular}{rrccrccrccrcc}
\hline
Star        & n &[La/Fe]{\sc ii}& rms & n & [Ce/Fe]{\sc ii} & rms &  n & [Eu/Fe]{\sc i}& rms \\ 
            &   & (dex)         &     &   &  (dex)          &     &    & (dex)         &     \\   
\hline
 77599 & 3& +0.44 &0.23 & 1 &$-$0.18 &     & 2&  +0.35 &0.29 \\ 
 83168 & 3& +0.43 &0.14 & 2 &$-$0.08 &0.31 & 2&  +0.34 &0.12 \\ 
101131 & 3& +0.44 &0.05 & 2 &$-$0.09 &0.26 & 2&  +0.21 &0.15 \\ 
108176 & 2& +0.41 &0.23 & 2 &$-$0.37 &0.35 & 2&  +0.15 &0.06 \\ 
108895 & 2& +0.17 &0.14 & 1 &$-$0.38 &     & 2&  +0.34 &0.15 \\ 
110677 & 3& +0.33 &0.12 & 1 &$-$0.49 &     & 2&  +0.28 &0.15 \\ 
111408 & 3& +0.28 &0.24 & 1 &$-$0.28 &     & 2&  +0.36 &0.19 \\ 
\hline
\end{tabular}
\label{t:ncaptu63a}
\end{table*}

\subsection{Comparison with literature}

To our knowledge, the only analysis of detailed elemental abundances in
NGC~6388 has been presented by Wallerstein et al. (2005). They used the
4m Blanco telescope at CTIO and observed 8 stars at R=35000
and high S/N (70 to 100). Their targets are much nearer the RGB tip than our 
sample (V=14.84 to 15.10 as compared to V=15.25 to 15.79).
They obtained an average [Fe/H]$\simeq-0.8$ and derived abundances for O and C
(found underabundant in comparison to most stars of similar metallicity, 
at [O/Fe]$\simeq-0.1$ [C/Fe]$\simeq-0.6$), Na and Al (slightly overabundant 
at [Na/Fe]$\simeq+0.3$, [Al/Fe]$\simeq+0.3$), the $\alpha$-elements Mg,
Si, Ca, Ti (defined about
normal, with [$\alpha$/Fe]$\simeq+0.15$),  and finally the $s$-process elements
Ba, La and the $r$-process one Ba showing small, normal excesses.

These results were presented so far only at a
conference and we miss details on the analysis. However, as the referee pointed
out, there may be some inconsistencies in their analysis, that would shed
doubts on the complete reliability of their results. In particular, when
corrected for the average cluster reddening, the magnitudes and colours
of the 8 stars appear to be matched by a [Fe/H]=$-0.4$ isochrone better than
by a [Fe/H]=$-0.8$ one. That implies a discrepancy both for temperatures
and gravities and a lower Fe abundance, i.e., altered elemental ratios.
We will not further discuss this analysis, awaiting for a more complete
presentation.

\begin{table*}
\caption[]{Mean abundance ratios for RGB stars in
the bulge clusters NGC~6388, NGC~6441, NGC~6528, NGC~6553 and in the Sgr cluster Terzan 7.
Whenever possible, values are referred to our solar reference abundances.}
\begin{tabular}{lcccccccccccc}
\hline
Ratio   & Mean& $\sigma$ &Mean& $\sigma$ &Mean& $\sigma$ &Mean& $\sigma$ &Mean & $\sigma$ &Mean & $\sigma$\\
        & NGC &          &NGC &          &NGC &          &NGC &          &Ter 7&          &Ter 7&         \\
        & 6388&          &6441&          &6528&          &6553&          &     &          &     &         \\
        &  (1)&   (1)    & (2)&   (2)    & (3)&   (3)    & (4)&   (4)    & (5) &  (5)     & (6) &   (6)   \\
\\
\hline
$[$O/Fe$]$I   & $-$0.30 & 0.16 &   +0.12 & 0.20 & $-$0.05 & 0.14 &   +0.38 & 0.13 &   +0.24 & 0.05 &         &     \\
$[$Na/Fe$]$I  &   +0.59 & 0.16 &   +0.55 & 0.15 &   +0.56 & 0.14 &         &	  & $-$0.14 & 0.10 &         &     \\
$[$Mg/Fe$]$I  &   +0.21 & 0.07 &   +0.34 & 0.09 &   +0.29 & 0.11 &   +0.56 & 0.10 &   +0.14 & 0.06 &   +0.08 & 0.07\\
$[$Al/Fe$]$I  &   +0.69 & 0.24 &   +0.30 & 0.15 &	  &	 &         &	  & $-$0.13 & 0.04 &         &     \\
$[$Si/Fe$]$I  &   +0.32 & 0.10 &   +0.33 & 0.11 &   +0.41 & 0.09 &   +0.19 & 0.18 &   +0.00 & 0.01 &   +0.13 & 0.09\\
$[$Ca/Fe$]$I  &   +0.06 & 0.06 &   +0.03 & 0.04 &   +0.16 & 0.06 &   +0.19 & 0.09 &   +0.06 & 0.03 &   +0.13 & 0.14\\
$[$Sc/Fe$]$II &   +0.05 & 0.06 &   +0.15 & 0.15 & $-$0.17 & 0.10 & $-$0.24 & 0.18 & $-$0.06 & 0.06 &         &     \\
$[$Ti/Fe$]$I  &   +0.37 & 0.10 &   +0.29 & 0.10 &   +0.05 & 0.08 &   +0.21 & 0.06 &   +0.13 & 0.03 &   +0.12 & 0.07\\
$[$Ti/Fe$]$II &   +0.30 & 0.12 &   +0.33 & 0.14 &	  &	 &	   &	  &   +0.11 & 0.02 &         &     \\
$[$V/Fe$]$I   &   +0.39 & 0.10 &   +0.29 & 0.14 & $-$0.14 & 0.09 &	   &	  &   +0.11 & 0.07 &         &     \\
$[$Cr/Fe$]$I  & $-$0.04 & 0.11 &   +0.15 & 0.18 & $-$0.01 & 0.05 &   +0.03 & 0.09 &   +0.02 & 0.03 &         &     \\
$[$Mn/Fe$]$I  & $-$0.25 & 0.02 &   +0.07 & 0.09 & $-$0.28 & 0.07 &         &      &   +0.00 & 0.05 &         &     \\
$[$Fe/H$]$I   & $-$0.44 & 0.04 & $-$0.43 & 0.06 &   +0.03 & 0.02 & $-$0.20 & 0.08 & $-$0.61 & 0.04 & $-$0.63 & 0.07\\
$[$Fe/H$]$II  & $-$0.37 & 0.09 & $-$0.42 & 0.24 &   +0.08 & 0.02 & $-$0.23 & 0.10 & $-$0.61 & 0.04 & $-$0.58 & 0.05\\
$[$Co/Fe$]$I  &   +0.04 & 0.07 &   +0.14 & 0.07 &	  &	 &	   &	  &   +0.01 & 0.08 &         &     \\
$[$Ni/Fe$]$I  &   +0.03 & 0.03 &   +0.13 & 0.07 &   +0.11 & 0.05 &   +0.02 & 0.07 & $-$0.11 & 0.02 & $-$0.19 & 0.05\\
$[$Y/Fe$]$I   & $-$0.16 & 0.19 & $-$0.05 & 0.24 &	  &	 &	   &	  &	    &	   &         &     \\
$[$Zr/Fe$]$I  & $-$0.16 & 0.12 & $-$0.40 & 0.19 &	  &	 &	   &	  & $-$0.07 & 0.07 &         &     \\
$[$Zr/Fe$]$II & $-$0.20 & 0.17 &	 &	&	  &	 &	   &	  &	    &	   &         &     \\
$[$Ba/Fe$]$II &   +0.21 & 0.10 &   +0.17 & 0.13 &   +0.17 & 0.08 &	   &	  &   +0.33 & 0.09 &         &     \\
$[$La/Fe$]$II &   +0.36 & 0.10 &	 &	&	  &	 &	   &	  &   +0.42 & 0.15 &         &     \\
$[$Ce/Fe$]$II & $-$0.27 & 0.16 &	 &	&	  &	 &	   &	  &   +0.22 & 0.07 &         &     \\
$[$Eu/Fe$]$II &   +0.29 & 0.08 &   +0.38 & 0.11 &	  &	 &	   &	  &   +0.53 & 0.05 &         &     \\
\hline
\end{tabular}

(1) present work, solar reference abundances from Gratton et al. (2003a)\\
(2) from Tab. 8 of Gratton et al. (Paper III), solar reference abundances from Gratton et al. (2003a)\\
(3) from Carretta et al. (2001), $after$ correction to present solar reference abundances\\
(4) from Cohen et al. (1999), $after$ correction to present solar reference abundances\\
(5) from Tautvaisiene et al. (2004), with unknown solar reference abundances\\
(6) from Sbordone et al. (2005), $after$ correction to present solar reference abundances 
\label{t:bulgegc}
\end{table*}

\section{Summary}

This is the first paper on the detailed chemical composition of the peculiar
bulge cluster NGC~6388, based on the analysis of high resolution UVES/FLAMES
spectra of seven RGB stars. 
From these stars (all cluster members) we find an average value 
[Fe/H]$=-0.44\pm0.01\pm0.03$ dex with no evidence of intrinsic spread in
metallicity in this cluster. We also obtained the detailed abundance pattern
for proton-capture elements (O, Na, Al, Mg), $\alpha-$elements (Si, Ca, Ti),
Fe-peak elements (Sc, V, Cr, Mn, Co, Ni) and neutron capture elements (Y, Zr,
Ba, La, Ce, Eu).

Studying the detailed chemistry of stars in such an old and moderately
metal-rich object opens to us a powerful window over the various kind of nucleosynthetic
sites that contribute to the formation and chemical enrichment of globular
clusters, allowing to isolate and study the yields provided by stars in
different ranges of masses and lifetimes.

The role of stars in the mass range $1 \le M \le 3-5 M_\odot$ is recognized
from the abundances of $s-$process elements like Ba, La, Ce: models for
chemical evolution in GCs must necessarily include nucleosynthesis from AGB
stars in this mass range.
However, within each single cluster the $s-process$ elements are homogeneusly
distributed (see James et al. 2004, Armosky et al. 1994): also in NGC~6388 
(and NGC~6441) the observed scatter is nearly
equal to or smaller than the spread expected from observational errors. 

These stars are not the same stars whose lifes and products affect the pattern of
proton-capture elements involved in H-burning at high temperatures (O, Na, Al,
Mg). The well known anti-correlation Na-O and Al-Mg established in NeNa and
MgAl cycles (and also found in NGC~6388) are not mirrored by similar variations
in $s-$process elements. The clear implication is that the abundance anomalies
are generated in stars of another (more large) mass range, likely intermediate 
mass AGB stars (4-8 $M_\odot$). 

We found in stars in NGC~6388 (as well as in the twin cluster NGC~6441)  an
excess of $\alpha-$process elements with respect to field disk and bulge stars
of similar metallicity, hint of a better capability to retain ejecta of massive
stars ending their life as core-collapse supernovae.
The same evidence is supported by another element typically produced in
massive stars environment: the pure $r-$process neutron capture element Eu. The
[Eu/Fe] ratios of NGC~6388 and NGC~6441 stand above the ratios found in field
stars and are rather similar to the value shown by low metallicity stars (both
in the field and in GCs). In the same way, the high [Eu/Y] values for these two
bulge clusters seem to be indicative of a dominance from $r-$process with
small contribution from the $s-$process (at least from the $main$ component).

In conclusion, it is still not easy to reconcile all the observational
facts into a coherent picture on how globular cluster formed and were enriched.
We have at hand contrasting evidences that the enrichment process is not
compatible with a pure self-enrichment scenario and, on the other hand, the
metal-rich GCs present proofs of enhanced efficiency in retain ejecta from the
massive SNe II with respect to field stars or looser open clusters.
More works, both on the theoretical and observational fronts, are needed before
a definitive conclusion might be reached.

\begin{acknowledgements}

We thank the anonymous referee for the very constructive comments.
This
publication makes use of data products from the Two Micron All Sky Survey,
which is a joint project of the University of Massachusetts and the Infrared
Processing and Analysis Center/California Institute of Technology, funded by
the National Aeronautics and Space Administration and the National Science
Foundation. This work was partially funded by the Italian MIUR
under PRIN 2003029437. We also acknowledge partial support from the grant
INAF 2005 ``Experimental nucleosynthesis in clean environments". 
\end{acknowledgements}


\begin{thebibliography}{}
\bibitem[]{}
 Alonso, A., Arribas, S. \& Martinez-Roger, C. 1999, A\&AS, 140, 261 

\bibitem[]{}
 Alonso, A., Arribas, S. \& Martinez-Roger, C. 2001, A\&A, 376, 1039 

\bibitem[]{}
 Arlandini, C., K\"appeler, F., Wisshak, K., Gallino, R., Lugaro, M., 
 Busso, M., Straniero, O. 1999, ApJ, 525, 886

\bibitem[]{}
 Armosky, B.J., Sneden, C., Langer, G.E., Kraft, R.P. 1994, AJ, 108, 1364

\bibitem[]{}
 Armandroff, T.E., Zinn, R. 1988, AJ, 96, 92
 
\bibitem[]{}
 Arnett, W.D. 1971, ApJ, 166, 153
 
\bibitem[]{}
 Bragaglia, A., Carretta, E., Gratton, R.G. et al. 2001, AJ, 121, 327

\bibitem[]{}
 Burris, D.L., Pilachowski, C.A., Armandroff, T.E., Sneden, C., Cowan, J.J.,
 Roe, H. 2000, ApJ, 544, 302

\bibitem[]{}
 Busso, G., Piotto, G., Cassisi, S. 2004, Mem.SAIt., 75, 46

\bibitem[]{}
 Cardelli, J.A., Clayton, G.C., \& Mathis, J.S. 1989, ApJ, 345, 245

\bibitem[]{} 
 Carretta, E. 2006, AJ, 131, 1766 


\bibitem[]{} 
 Carretta, E., Gratton, R.G. 2006, in preparation 

\bibitem[]{}
 Carretta, E., Cohen, J., Gratton, R.G. \& Behr, B.B. 2001, AJ, 122, 1469

\bibitem[]{} 
 Carretta, E., Gratton R.G., Bragaglia, A., Bonifacio, P. \&  Pasquini, L. 
 2004, A\&A, 416, 925 


\bibitem[]{}
 Carretta, E., Bragaglia, A., Gratton R.G., Leone, F., Recio-Blanco,
 A., Lucatello, S. 2006a, A\&A, 450, 523 (Paper I) 

\bibitem[]{}
 Carretta, E., Bragaglia, A., Gratton R.G., Lucatello, S., \&
 Momany, Y. 2006b, A\&A, in press  (Paper II) 

\bibitem[]{}
 Carretta et al. (2006c), A\&A, in press (Paper IV) 

\bibitem[]{}
 Carretta et al. (2006d), in preparation

\bibitem[]{} 
 Cayrel, R. 1986, A\&A, 168, 8

\bibitem[]{}
 Cohen, J.G., Gratton, R.G., Behr, B.B., Carretta, E. 1999, ApJ, 523, 739

\bibitem[]{}
 Corwin, T.M., Catelan, M., Borissova, J., Smith, H.A. 2004, A\&A, 421, 667

\bibitem[]{}
 Corwin, T.M., Sumerel, A.N, Pritzl, B.J., Smith, H.A., Catelan, M., Sweigart,
 A.V., Stetson, P.B. 2006, AJ, in press (astro-ph/0605569)

\bibitem[]{} 
 D'Antona, F. \& Caloi, V. 2004, ApJ, 611, 871


\bibitem[]{}
 Feltzing, S. \& Gustafsson, B. 1998, A\&AS, 129, 237
 
\bibitem[]{}
 Gratton, R.G. 1988, Rome Obs. Preprint, 29

\bibitem[]{}
 Gratton, R.G. 1989, A\&A, 208, 171

\bibitem[]{}
 Gratton, R.G., Carretta, E., Eriksson, K., \& Gustafsson, B. 1999,
 A\&A, 350, 955 

\bibitem[]{}
 Gratton, R.G., Carretta, E., Claudi, R., Lucatello, S., \&
 Barbieri, M. 2003a, A\&A, 404, 187 

\bibitem[]{}
 Gratton, R.G., Carretta, E., Desidera, S., Lucatello, S., Mazzei, P. \&
 Barbieri, M. 2003b, A\&A, 406, 131 

\bibitem[]{} 
 Gratton, R.G., Sneden, C., \& Carretta, E. 2004, ARA\&A, 42, 385

\bibitem[]{} 
 Gratton, R.G., Bragaglia, A., Carretta, E., Lucatello, S., Momany, Y.,
 Pancino, E., \& Valenti, E. 2006, A\&A, in press (Paper III) 

\bibitem[]{} 
 Gratton, R.G. et al. 2007, A\&A, in press (Paper V) 

\bibitem[]{}
 Harris, W.~E. 1996, AJ, 112, 1487 

\bibitem[]{}
 James, G., Fran\c cois, P., Bonifacio, P., Carretta, E., Gratton, R.G., 
 Spite, F. 2004, A\&A, 427, 825

\bibitem[]{} 
 Kurucz, R.L. 1993, CD-ROM 13, Smithsonian Astrophysical

\bibitem[]{}
 Lynch, D.K., Bowers, P.F., Whiteoak, J.B. 1989, AJ, 97, 1708

\bibitem[]{}
 Magain, P. 1984, A\&A, 134, 189

\bibitem[]{}
 McLaughlin, D.E., van der Marel, R.P. 2005, ApJ.Suppl. Ser., 161, 304

\bibitem[]{}
 McWilliam, A., Preston, G.W., Sneden, C., Searle, L. 1995, AJ, 109, 2757

\bibitem[]{}
 McWilliam, A., Rich, R.~M., Smecker-Hane, T.~A. 2003, ApJ, 592, L21  

\bibitem[]{}
 Moehler, S., Sweigart, A.V. 2006a, Baltic Astronomy, 15,41

\bibitem[]{}
 Moehler, S., Sweigart, A.V. 2006b, A\&A, in press, astro-ph/0606054

\bibitem[]{}
 Moehler, S., Sweigart, A.V., Catelan, M. 1999, A\&A, 351, 519

\bibitem[]{} 
 Momany, Y., Held, E.~V., Saviane, I., Rizzi, L. 2002, A\&A, 384, 393 
 
\bibitem[]{}
 Momany, Y., Cassisi, S., Piotto, G., Bedin, L.R., Ortolani, S., Castelli, F., 
 Recio-Blanco, A. 2003, A\&A, 407, 303 

\bibitem[]{}
 Pasquini, L. et al. 2002, The Messenger, 110, 1

\bibitem[]{}
 Pritzl, B.J., Smith, H.A., Catelan, M., Sweigart, A.V. 2001, AJ, 122, 2600

\bibitem[]{}
 Pritzl, B.J., Smith, H.A., Catelan, M., Sweigart, A.V. 2002, AJ, 124, 949

\bibitem[]{}
 Pritzl, B.J., Venn, K.A., Irwin, M.J. 2005, AJ, 130, 2140
 
\bibitem[]{}
 Raimondo, G., Castellani, V., Cassisi, S., Brocato, E., Piotto, G. 2002, ApJ,
 569, 975

\bibitem[]{}
 Reddy, B.E., Tomkin, J., Lambert, D.L., Allende prieto, C. 2003, MNRAS, 340,
 304
 
\bibitem[]{}
 Rich, R.M. et al. 1997, ApJ, 484, L25

\bibitem[]{}
 Ryan, S.G., Norris, J.E., beers, T.C. 1996, ApJ, 471, 254

\bibitem[]{}
 Sandage, A. 2006, AJ, 131, 1750
  

\bibitem[]{} 
 Skrutskie, M.F. et al. 2006, AJ, 131, 1163

\bibitem[]{}
 Sobeck, J.~S., Ivans, I.~I., Simmerer, J.~A., Sneden, C., Hoeflich, P., 
 Fulbright, J.~P., Kraft, R.~P. 2006, AJ, 131, 2949 

\bibitem[]{}
 Sneden, C., Kraft, R.P., Shetrone, M.D., Smith, G.H., Langer, G.E., Prosser,
 C.F. 1997, AJ, 114, 1964

\bibitem[]{}
 Sneden, C., Pilachowski, C., Kraft, R.P. 2000, AJ, 120, 1351

\bibitem[]{}
 Stetson, P.~B. 1994, PASP, 106, 250

\bibitem[]{}
 Sweigart, A.V., Catelan, M. 1998, ApJ, 501, L63

\bibitem[]{}
 Tautvaisiene, G., Wallerstein, G., Geisler, D., Gonzalez, G. \& Charbonnel, C.
 2004, AJ, 127, 373

\bibitem[]{}
 Thoul, A., Jorissen, A., Goriely, S., Jehin, E., Magain, P. Noels, A., 
 Parmentier, G. 2002, A\&A, 383, 491

\bibitem[]{}
 Travaglio, C., Gallino, R., Arnone, E., Cowan, J, Jordan, F., Sneden, C. 2004,
 ApJ, 601, 864
 
\bibitem[]{}
 Truran, J. 1988, in The Impact of Very High S/N Spectroscopy on Stellar Physics,
 eds. G. Cayrel de Strobel, M. Spite,  IAU Symp. 132, 577

\bibitem[]{}
 Truran, J. 1991, ASP Conf. Ser. 13, 78

\bibitem[]{}
 Venn, K.A., Irwin, M., Shetrone, M.D., Tout, C.A., Hill, V., Tolstoy, E. 2004,
 AJ, 128, 1177

\bibitem[]{} 
 Wallerstein, G., Kovtyukh, V., \&  Andrievsky, S. 2005, in From Lithium to 
 Uranium: Elemental Tracers of Early Cosmic Evolution, eds. V. Hill, P.
 Fran\c cois, F.Primas, IAU Symp. 228, 413

\bibitem[]{}
 Yi, S., Demarque, P., Oemler, A. 1998, ApJ, 492, 480

\bibitem[]{} 
 Zacharias, N., Urban, S.~E., Zacharias, M.~I., Wycoff, G.~L., 
 Hall, D.~M., Monet, D.~G., Rafferty, T.~J. 2004, AJ, 127, 3043

\end{thebibliography}
\end{document}